\pdfoutput=1
\documentclass[sn-mathphys-num]{sn-jnl}
\usepackage{graphicx}
\usepackage{multirow}%
\usepackage{amsmath,amssymb,amsfonts}%
\usepackage{amsthm}%
\usepackage{mathrsfs}%
\usepackage[title]{appendix}%
\usepackage{xcolor}%
\usepackage{textcomp}%
\usepackage{manyfoot}%
\usepackage{booktabs}%
\usepackage{algorithm}%
\usepackage{algorithmicx}%
\usepackage{algpseudocode}%
\usepackage{listings}%

\usepackage{epsfig}
\usepackage{subcaption}
\usepackage{hyperref}
\usepackage{tabularx}
\usepackage{float}
\usepackage{pifont}
\usepackage{tikz}
\usepackage{pgfplots}
\pgfplotsset{compat=1.18}
\usetikzlibrary{positioning}

\usepackage{xcolor}
\usepackage[normalem]{ulem} 

\newcommand{\added}[1]{#1}
\newcommand{\deleted}[1]{}

\theoremstyle{thmstyleone}%
\theoremstyle{thmstyletwo}%

\theoremstyle{thmstylethree}%

\begin{document}

\title[A Survey of Threats Against Voice Authentication and Anti-Spoofing Systems]{A Survey of Threats Against Voice Authentication and Anti-Spoofing Systems}

\author*[a]{\fnm{Kamel} \sur{Kamel}}\email{kamel.kamel@research.deakin.edu.au}

\author[b]{\fnm{Keshav} \sur{Sood}}\email{keshav.sood@deakin.edu.au}

\author[c]{\fnm{Hridoy Sankar} \sur{Dutta}}\email{hridoy.dutta@deakin.edu.au}

\author[a]{\fnm{Sunil} \sur{Aryal}}\email{sunil.aryal@deakin.edu.au}

\affil*[a]{\orgdiv{School of Information Technology}, \orgname{Deakin University}, \orgaddress{\street{75 Pigdons Road}, \city{Waurn Ponds}, \postcode{3216}, \state{Victoria}, \country{Australia}}}

\affil[b]{\orgdiv{School of Information Technology}, \orgname{Deakin University, Burwood Campus}, \orgaddress{\street{221 Burwood Highway}, \city{Burwood}, \postcode{3125}, \state{Victoria}, \country{Australia}}}

\affil[c]{\orgdiv{School of Information Technology}, \orgname{Deakin University, GIFT City Campus}, \orgaddress{\street{GIFT City}, \city{Gandhinagar}, \postcode{382355}, \state{Gujarat}, \country{India}}}

\abstract{
Voice authentication has evolved from traditional systems based on hand-engineered acoustic features to deep learning models that learn speaker representations directly from audio. This progress has improved accuracy, but it has also opened new attack surfaces. Prior surveys treat these threats in isolation; we instead provide a unified analysis of the four primary attack vectors against voice authentication: data poisoning, adversarial perturbations, audio deepfakes, and adversarial spoofing. We organize them along three axes: what the attacker knows, how the attack is delivered, and how broadly it generalizes across speakers and inputs. Comparing these vectors reveals a clear gap in threat maturity. Adversarial perturbations are effective when the attacker has full model access, but often fail once the audio is played over the air. Deepfakes, by contrast, scale easily, succeed without model access, and routinely fool both machines and human listeners. Existing Anti-Spoofing Countermeasures (CMs) remain largely reactive and can themselves be bypassed by adaptive attacks. We close by mapping open challenges and arguing for standardized threat models and defenses that act proactively, in real time.}


\maketitle

\section{Introduction}

Voice authentication is a biometric verification method that leverages the unique characteristics of an individual's voice to authenticate their identity. It originated as a simple speaker recognition technique in the mid-20th century, relying primarily on handcrafted acoustic features and statistical models such as Gaussian Mixture Models (GMMs). It was initially adopted in restricted environments like forensic analysis and limited-access security systems, where its development was constrained by speech condition variation and limited computational capabilities. However, the landscape changed dramatically with the emergence of deep learning. Over the past decade, deep learning has fundamentally transformed voice authentication, enabling more accurate, robust, and scalable solutions. Advanced models, such as Deep Neural Networks (DNNs), Convolutional Neural Networks (CNNs), and Transformer-based architectures, now extract complex and invariant voice representations, often referred to as speaker embeddings, which significantly improve performance in real-world environments ~\cite{o2023review, sharma2024milestones, jakubec2024deep, hanifa2021review}.

Voice authentication plays a critical role in modern security ecosystems due to its non-intrusive nature, ease of implementation, and cost-effectiveness. Its applications span a wide range of domains, from mobile device unlocking to voice assistants, secure banking transactions, smart home systems, and access control in law enforcement and judicial processes \cite{singh2012applications, algabri2017automatic, rose2006technical}. As a contactless solution requiring only a microphone, voice authentication enhances user experience while mitigating risks associated with password leakage, social engineering, and physical spoofing, making it an increasingly important component of multi-factor authentication frameworks \cite{markowitz2000voice}.

The first documented attack on a voice authentication system was recorded in the 1990s when researchers discovered that replaying a pre-recorded voice sample could effectively bypass early speaker verification systems \cite{lindberg1999vulnerabilities}. These systems lacked robust liveness detection and could not distinguish between a real speaker and a recorded signal, exposing a fundamental weakness that undermined their security in practical scenarios. This initial form of the replay attack became a foundational example in the study of biometric spoofing and highlighted the necessity of building systems capable of discerning not just who is speaking, but also how and when the speech occurs \cite{villalba2011detecting, al2025convolutional}.

The sophistication of these attacks has since escalated through successive phases. The 2010s saw the emergence of machine-learning-based voice synthesis, transforming voice conversion from a signal-processing curiosity into a practical spoofing tool. The mid-2010s brought gradient-based adversarial attacks, enabling imperceptible manipulations of raw audio waveforms that could reliably deceive deep neural network–based verifiers \cite{helpnetsecurity2010voiceEncryptionVulnerability, assessment2016}. By the early 2020s, generative diffusion models and zero-shot voice cloning had democratized high-fidelity voice synthesis, lowering the cost of an impersonation attack from hours of recording to seconds. This escalation culminated in multimodal deepfake incidents where synchronized audio and video forgeries are used to deceive victims in real time \cite{darkreading2020uae}.

Beyond the technical progression of attack methods, recent real-world incidents show that these threats have already moved from laboratory demonstrations to operational fraud. In 2019, fraudsters successfully cloned the voice of a German CEO to deceive a UK energy executive into transferring €220,000 to a fraudulent supplier \cite{avast2019deepfake}, marking one of the first widely publicized AI voice synthesis attacks. The threat escalated further in 2020, when criminals used AI voice synthesis combined with email spoofing to impersonate a company director and defraud a UAE bank of approximately \$35 million \cite{darkreading2020uae, brewster2021fraudsters, tran2021bank}. Most recently, in 2024, a finance employee at Arup Engineering was deceived into transferring \$25 million during a live video call featuring AI-generated deepfakes of multiple company executives \cite{arup2024deepfake}, illustrating that attackers have progressed from simple audio-only impersonation to sophisticated multi-modal deception. These incidents collectively demonstrate that voice authentication vulnerabilities are no longer theoretical concerns but active vectors for large-scale financial fraud, underscoring the urgent need for comprehensive security measures.

This incident highlights a growing concern: as voice authentication systems become more sophisticated, so do the methods used to compromise them. One critical threat is \textbf{data poisoning}, where adversaries inject malicious or mislabeled data into training datasets or pipelines to compromise model behavior or insert hidden backdoors \cite{DPsurvey, jagielski2018manipulating, biggio2012poisoning, ge2023data, yan2024backdoor}. While extensively studied in Computer Vision and NLP, poisoning remains underexplored in the audio domain, lacking standardized benchmarks and threat models. Another challenge comes from \textbf{adversarial attacks}, which involve subtle audio perturbations that trick models into misclassification or unauthorized authentication. Though some defenses exist, they often fail under realistic deployment conditions, such as black-box or over-the-air scenarios \cite{xu2020adversarial, machado2021adversarial, lan2022adversarial, chen2021real}.

Meanwhile, \textbf{deepfake voice attacks} are evolving very quickly. High-fidelity synthetic voices, generated from minimal samples, are now capable of effectively mimicking real users, making it easier to bypass voice authentication systems \cite{wu2020voice, kinnunen2020tandem}. Existing countermeasures, including replay and spoofing detectors, are increasingly outpaced by these generative techniques. Even the defensive modules themselves have become targets. Attackers now design \textbf{adversarial spoofing attacks} to evade VAS and spoofing detectors, undermining the reliability of traditional protection measures. This has led to a security arms race in which new defenses trigger more advanced attack strategies, emphasizing the need for holistic, adaptive, and robust approaches to secure voice authentication.

\added{Despite this surge of interest, research on attacks against voice authentication remains fragmented: data poisoning, adversarial perturbations, deepfake synthesis, and anti-spoofing evasion are each studied---and surveyed---in isolation, leaving practitioners without a single reference that relates these vectors, compares their feasibility, and reveals how they compound. The central aim of this survey is to close that gap.} In this paper, we provide a comprehensive survey of the threats targeting voice authentication systems. These threat channels exploit vulnerabilities across the machine learning pipeline, from model training to input inference and security filtering, and challenge the reliability of speaker verification technology. By synthesizing findings from recent literature, we present a structured overview of how these threats operate, how current systems respond, and where critical security gaps remain. Our contributions are summarized below:

\begin{enumerate}
    \item \textbf{Taxonomy \& Categorization}: We propose a comprehensive taxonomy that categorizes threats based on critical dimensions, including attacker knowledge, delivery channels, and perceptual universality, to provide a structured understanding of the vulnerability landscape.
    \item \textbf{A Survey of Attack Methodologies}: For each attack category, we summarize current methodologies, highlight commonly used datasets, and compare their performance and limitations using widely accepted taxonomies.
    
    \item \textbf{Identification of Future Directions}: By highlighting emerging risks and open challenges, this survey aims to support the development of more secure and resilient voice authentication systems.
\end{enumerate}
While prior surveys have explored specific attack classes, such as backdoor attacks \cite{yan2024backdoor}, adversarial attacks \cite{lan2022adversarial}, and audio deepfake detection \cite{yi2023audio, zhang2025audio, almutairi2022review}, they have overlooked key areas like broader data poisoning techniques, deepfake attacks themselves, and adversarial spoofing attacks. Moreover, there is currently no unified examination that addresses the full spectrum of attacks facing voice authentication systems. Existing works generally treat these threats in isolation, overlooking the interplay and compounding risks that arise when multiple attack strategies are combined. This survey fills that gap by providing the first unified treatment that addresses all four threat vectors -- data poisoning, adversarial, deepfake attacks, and adversarial spoofing attacks -- within a single taxonomic framework, enabling systematic cross-threat comparison. \added{Table~\ref{tab:survey_comparison} contrasts the coverage of this survey against the most closely related prior surveys, none of which spans all four threat vectors or treats adversarial spoofing (attacks on the countermeasures themselves).}

\begin{table*}[t]
\centering
\caption{\added{Coverage of the most closely related surveys versus this survey across the four voice-authentication attack families (\ding{51}: substantive coverage; --: little or no treatment).}}
\label{tab:survey_comparison}
\added{\resizebox{\textwidth}{!}{%
\begin{tabular}{@{}lccccc@{}}
\toprule
\textbf{Survey} & \textbf{Data Poisoning} & \textbf{Adversarial} & \textbf{Deepfake} & \textbf{Adv.\ Spoofing} & \textbf{Unified Taxonomy} \\
\midrule
Yan et al.~\cite{yan2024backdoor} & \ding{51} & -- & -- & -- & -- \\
Lan et al.~\cite{lan2022adversarial} & -- & \ding{51} & -- & -- & -- \\
Tan et al.~\cite{tan2022adversarial} & -- & \ding{51} & -- & -- & -- \\
Yi et al.~\cite{yi2023audio} & -- & -- & \ding{51} & -- & -- \\
Almutairi et al.~\cite{almutairi2022review} & -- & -- & \ding{51} & -- & -- \\
Zhang et al.~\cite{zhang2025audio} & -- & -- & \ding{51} & -- & -- \\
\textbf{This survey} & \ding{51} & \ding{51} & \ding{51} & \ding{51} & \ding{51} \\
\bottomrule
\end{tabular}}}
\end{table*}

\added{\textbf{Scope and limitations.} The objective of this survey is to systematically characterize the attack surface of voice authentication systems across the four threat vectors above, so that researchers and practitioners can reason about the full landscape of risks within a single framework. To keep this objective tractable, two related directions are deliberately placed out of scope. First, we do not conduct a unified empirical re-benchmarking of attacks under a common experimental protocol; the results we report are drawn from the heterogeneous datasets, models, and metrics used in the primary studies, and a controlled re-evaluation would constitute a distinct contribution in its own right. Second, although we summarize representative countermeasures where they clarify the significance of an attack, we do not provide a systematic treatment of defenses, which form a large and rapidly evolving literature that merits a dedicated review. Finally, our analysis targets attacks at the model and signal level rather than protections at the network or transport layer; the latter are complementary but orthogonal to the biometric threat surface examined here.}

This paper is organized as follows. Section \ref{sec:methodology} describes the methodology used to select and analyze papers for this survey, including our inclusion and exclusion criteria. Section \ref{sec:background} provides background on voice authentication systems. Section \ref{sec:taxonomies_and_threats} presents taxonomies and the typical threat models these systems face. Section \ref{sec:data_poisoning} examines data poisoning attacks, discussing techniques that compromise training data to subvert model behavior. Section \ref{sec:adv_attacks} delves into adversarial attacks, where carefully crafted audio inputs are designed to mislead voice authentication systems during inference. Section \ref{sec:deepfake_attacks} investigates deepfake attacks, focusing on the creation of synthetic voices and their potential to impersonate legitimate users. Section \ref{sec:adv_attacks_cms} addresses adversarial spoofing attacks, which target anti-spoofing countermeasures to allow malicious audio to bypass security layers. Section \ref{sec:evaluation} surveys prominent datasets, benchmarks, and competitions that support research and evaluation in this domain. Section \ref{sec:realworld} discusses real-world deployments and identifies existing gaps in current implementations. Section \ref{sec:comparison} offers a comparison of the different attack strategies discussed. Section \ref{sec:future_work} outlines directions for future work, suggesting ways to enhance system robustness. Finally, Section \ref{sec:conculsion} provides a conclusion, summarizing the key insights and implications of this survey.
\section{Methodology}
\label{sec:methodology}
To ensure a comprehensive and unbiased survey of the threat landscape, we employed a systematic screening methodology.

\subsection{Search Strategy and Data Sources}
We conducted a structured search across major academic databases, including IEEE Xplore, ACM Digital Library, SpringerLink, ScienceDirect, and arXiv. The search covered publications from January 2010 to March 2026, spanning the emergence of deep learning-based voice authentication systems through the most recent advances in generative and adversarial threats.

The search queries were constructed using Boolean operators, combining terms from two clusters:

\begin{itemize}
    \item \textbf{Target System:} ``Voice Authentication" 
    OR ``Speaker Verification" 
    OR ``Speaker Recognition" 
    OR ``Anti-spoofing".
    \item \textbf{Attack Type:} ``Adversarial Attack" 
    OR ``Data Poisoning" 
    OR ``Backdoor" 
    OR ``Deepfake" 
    OR ``Voice Conversion" 
    OR ``Spoofing".
\end{itemize}

The full combined query applied across all databases was:

\begin{quote}
\textit{(``Voice Authentication'' OR ``Speaker Verification'' OR ``Speaker Recognition'' OR ``Anti-spoofing'') AND (``Adversarial Attack'' OR ``Data Poisoning'' OR ``Backdoor'' OR ``Deepfake'' OR ``Voice Conversion'' OR ``Spoofing'')}
\end{quote}

Minor syntactic adaptations were applied per database (e.g., field tags \texttt{abs:} on arXiv, query split into two sub-queries of eight fields on ScienceDirect due to platform constraints), while preserving the logical structure of the query.

\subsection{Inclusion and Exclusion Criteria}
We filtered the initial search results based on the following criteria to ensure relevance and quality:

\begin{itemize}
    \item \textbf{Inclusion:} Peer-reviewed conference papers and journal articles that propose a novel attack against VAS; studies providing empirical evaluations of existing threats; and high-impact preprints on arXiv from top-tier venues.
    \item \textbf{Exclusion:} Papers focusing solely on Automatic Speech Recognition (ASR) without a biometric component; studies on off-topic spoofing domains (e.g., face anti-spoofing, GPS/GNSS spoofing, network     IP spoofing); non-technical articles, tutorials, or duplicates; and studies lacking quantitative evaluation metrics (e.g., EER, ASR).
\end{itemize}

\subsection*{Search Results and Screening}
The database search retrieved a total of 4,957 records (IEEE Xplore: 1,694; ScienceDirect: 1,077; ACM Digital Library: 823; 
SpringerLink: 776; arXiv: 587). After removing 325 cross-database duplicates, 4,632 unique records were screened by title. Of these, 
2,676 were excluded as off-topic (primarily face anti-spoofing, GPS spoofing, and network-layer spoofing), leaving 1,956 candidates for 
full-text eligibility assessment. Following full-text review against the inclusion and exclusion criteria, 86 studies were finally included 
in this survey. The complete screening process is illustrated in Figure~\ref{fig:prisma}. \added{We further note that, where preprints are included, they correspond predominantly to recent, high-impact contributions that are either already widely cited by the community or have since appeared in peer-reviewed venues; non-peer-reviewed sources are retained only for very recent developments for which no published version is yet available.}

\begin{figure}[h]
\centering
\begin{tikzpicture}[
  box/.style={rectangle, draw, rounded corners=4pt, 
              text width=6.5cm, align=center, 
              minimum height=1.1cm, font=\small},
  exc/.style={rectangle, draw, dashed, rounded corners=4pt, 
              text width=4.5cm, align=center, 
              minimum height=1.1cm, font=\small, fill=gray!8},
  arr/.style={->, thick},
  side/.style={->, thick, dashed}
]

\node[box, fill=blue!6]  (id)  
      {\textbf{Identification}\\[2pt] Records retrieved from 5 databases\\$n = 4{,}957$};

\node[box, fill=blue!6, below=1.0cm of id]  (dedup)  
      {\textbf{After Deduplication}\\[2pt] Unique records screened\\$n = 4{,}632$};

\node[exc, right=1.2cm of dedup] (dup)  
      {Duplicates removed\\$n = 325$};

\node[box, fill=yellow!10, below=1.0cm of dedup] (screen)  
      {\textbf{Title Screening}\\[2pt] Records after off-topic exclusion\\$n = 1{,}956$};

\node[exc, right=1.2cm of screen] (excl)  
      {Excluded (off-topic)\\$n = 2{,}676$};

\node[box, fill=green!8, below=1.0cm of screen]  (included)  
      {\textbf{Finally Included}\\[2pt] Studies included in the survey\\$n = 86$};

\node[exc, right=1.2cm of included] (excl2)  
      {Excluded after\\full-text review\\$n =$ \deleted{1{,}874}\added{1{,}870}};

\draw[arr] (id)     -- (dedup);
\draw[arr] (dedup)  -- (screen);
\draw[arr] (screen) -- (included);
\draw[side] (dedup)   -- (dup);
\draw[side] (screen)  -- (excl);
\draw[side] (included)-- (excl2);
\end{tikzpicture}
\caption{PRISMA-style flow diagram of the systematic search and 
screening process.}
\label{fig:prisma}
\end{figure}

\section{Background}
\label{sec:background}

Attacks on voice authentication only make sense against the architecture they target. This section reviews how modern Voice Authentication Systems (VAS) are built, the verification tasks they perform, and the frameworks used to categorize the threats they face.

\begin{table*}[htbp]
\centering
\caption{List of Acronyms}
\label{tab:acronyms}
\resizebox{\textwidth}{!}{
\begin{tabular}{ll|ll}
\toprule
\textbf{Acronym} & \textbf{Definition} & \textbf{Acronym} & \textbf{Definition} \\ \midrule
SV   & Speaker Verification            & VAS  & Voice Authentication Systems \\
ASV  & Automatic Speaker Verification  & VC   & Voice Conversion             \\
SRS  & Speaker Recognition System      & TTS  & Text-to-Speech               \\
SRE  & Speaker Recognition Evaluations & ADD  & Audio Deepfake Detection     \\
CSI  & Closed-set Identification       & DFDC & Deepfake Detection Challenge \\
OSI  & Open-set Identification         & PAD  & Presentation Attack Detection \\ \midrule
PF   & Partial Fake                    & GMMs & Gaussian Mixture Models      \\
CMs  & Anti-spoofing Countermeasures   & UBM  & Universal Background Model   \\
DNN  & Deep Neural Network             & CNN  & Convolutional Neural Networks \\
k-NN & k-Nearest Neighbors             & GAN  & Generative Adversarial Network \\ \midrule
MFCC & Mel-Frequency Cepstral Coefficients & GFCC & Gammatone Frequency Cepstral Coefficients \\
SNR  & Signal-to-Noise Ratio           & RIR  & Room Impulse Response        \\
FIR  & Finite Impulse Response         & DET  & Detection Error Tradeoff     \\
EER  & Equal Error Rate                & ASR  & Attack Success Rate          \\ \midrule
FGSM & Fast Gradient Sign Method       & BIM  & Basic Iterative Method       \\
PGD  & Projected Gradient Descent      & NES  & Natural Evolution Strategy   \\
UAPs & Universal Adversarial Perturbations &      &                              \\ \bottomrule
\end{tabular}
}
\end{table*}
\subsection{Voice Authentication Systems Overview}
Voice authentication systems have evolved significantly over the years, transitioning from traditional statistical models to advanced deep learning-based architectures. Understanding the architecture and taxonomy of these systems is essential for identifying their vulnerabilities, operational contexts, and security challenges. This section provides a detailed overview of the underlying architectures and the different types of voice authentication tasks.

\subsubsection{System Architectures}
\begin{figure*}[h]
    \centering
    \includegraphics[width=\textwidth]{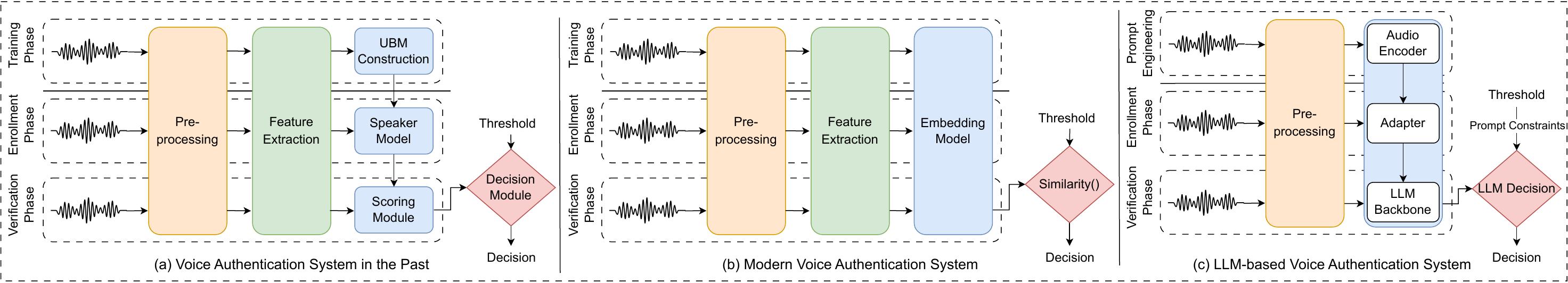}
    \caption{Voice Authentication Systems: (a) Traditional feature-based models. (b) Modern deep learning–based systems. (c) LLM-based systems combining acoustic and linguistic cues for context-aware authentication.}
    \label{fig:voice_auth_comparison}
\end{figure*}

A typical speaker recognition system (SRS) operates in three main stages, as illustrated in Figure~\ref{fig:voice_auth_comparison}. In the \textbf{training phase}, the system learns from a large dataset of voices to extract speakers' discriminative features and build a robust embedding space, often using deep learning models. During the \textbf{enrollment phase}, each user's voiceprint is captured through multiple samples and transformed into a reference embedding that is securely stored. Finally, in the \textbf{recognition phase}, a newly captured voice sample is embedded and compared against the enrolled reference using similarity scoring and thresholding strategies.

In traditional systems (Figure~\ref{fig:voice_auth_comparison}.a), the process begins with pre-processing and feature extraction, followed by training a Universal Background Model (UBM)\cite{reynolds2000speaker}. Speaker-specific models are derived from the UBM during enrollment, and verification is performed using statistical techniques such as Gaussian Mixture Models (GMMs)\cite{reynolds1995robust}. Models like i-vector \cite{ivector} significantly improved system compactness and efficiency by encoding speaker characteristics into low-dimensional embeddings. However, this multi-stage architecture is sensitive to environmental noise and variability in speaker conditions.

Modern systems (Figure~\ref{fig:voice_auth_comparison}.b) have adopted end-to-end deep learning architectures. These systems unify pre-processing, feature extraction, and speaker modeling into a single pipeline. Rather than constructing explicit speaker models, embeddings are extracted directly from speech signals and compared using similarity metrics like cosine similarity or learned distance functions \cite{xvectors, deepspeaker}. While these models offer improved accuracy and scalability, they still face challenges such as environmental noise, microphone variability, and resistance to spoofing attacks (e.g., replay, synthesis, and voice conversion).

Emerging LLM-based approaches \cite{ren2025can, mitrofanov2025cryfish} (Figure~\ref{fig:voice_auth_comparison}.c) mark a conceptual shift in speaker verification by reframing it as a multimodal question-answering or generative task. Rather than relying exclusively on embedding similarity, these systems, commonly referred to as Audio Large Language Models (AudioLLMs), couple pre-trained audio encoders (e.g., WavLM \cite{chen2022wavlm}) with Large Language Models (e.g., Qwen2 \cite{wang2024qwen2}) through lightweight adapter layers. This design enables joint reasoning over acoustic representations and textual prompts, allowing the model to produce natural-language judgments about speaker identity. Although early zero-shot evaluations showed that such general-purpose models underperformed compared to specialized architectures such as ECAPA-TDNN \cite{desplanques2020ecapa} on conventional verification benchmarks, recent progress in instruction tuning, particularly through hard pair sampling strategies, has substantially narrowed this gap. In some short-utterance settings, these models now match or even surpass traditional approaches. Moreover, their semantic reasoning capabilities naturally support unified text-dependent verification, in which both the speaker's identity and the correctness of the spoken content are assessed within a single framework.

From a security standpoint, LLM-based VAS introduce a qualitatively distinct threat surface that does not map cleanly onto the classical taxonomy developed for embedding-based systems. Because these systems process joint acoustic and textual representations, they are susceptible to \textbf{audio-based prompt injection}: an attacker could embed adversarially crafted speech commands within an audio stream that redirect the model's natural-language reasoning while appearing benign to human listeners\deleted{ \mbox{\cite{karaouglan2024novel}}}. Similarly, the language-model component of AudioLLMs opens a pathway for \textbf{jailbreaks via synthesized speech}, where carefully constructed prosodic or lexical patterns bypass safety guardrails encoded at the language-model level. These threats represent an emerging frontier that the adversarial audio community has only begun to characterize, and are discussed further in Section~\ref{sec:future_work}.

\subsubsection{Voice Authentication Tasks}
Voice authentication tasks can be categorized along three primary dimensions: the recognition task type, the interaction mode, and the content dependency. Each dimension presents distinct technical challenges and application scenarios.

\textbf{Recognition task types} range from one-to-one \textit{Speaker Verification (SV)}, which accepts a claimed identity if a similarity score exceeds a threshold, to one-to-many \textit{Closed-set Identification (CSI)}, which always assigns an identity from a known set, and \textit{Open-set Identification (OSI)}, which extends CSI with a rejection option for unrecognized speakers. \textit{Speaker Diarization} partitions an audio stream into speaker-labelled segments without prior enrollment. SV is the most security-critical task because it is the direct target of impersonation attacks; CSI is uniquely vulnerable because its lack of rejection forces assignment even for impostor inputs.

\textbf{Interaction modes} determine how much attacker control over the acoustic environment is possible. \textit{Active systems} require explicit prompted speech (passphrases, challenge-response) and constrain acoustic conditions, reducing replay risk at the cost of user friction. \textit{Passive systems} authenticate continuously during natural conversation, offering a seamless user experience but exposing a larger acoustic surface to interception and synthesis attacks.

\textbf{Content dependency} affects which features the model relies on. \textit{Text-dependent systems} bind authentication to fixed phrases, making them vulnerable to replay of those specific phrases but simpler to protect with dynamic challenge prompts. \textit{Text-independent systems} accept arbitrary content, requiring the model to generalize speaker identity across phonetic contexts, a richer but harder-to-harden representation that is the primary target of adversarial perturbation attacks.

Modern deployed systems often combine these modes, such as passive text-independent verification during routine interactions with active text-dependent challenge for high-value transactions.

\subsection{Attacks Definitions in Voice Authentication}
Despite significant advances, voice authentication systems remain vulnerable to a diverse range of attacks that target both the training and inference phases. These attacks are often motivated by the goal of impersonating users, bypassing authentication, or degrading system performance.

To better understand the evolving focus of the research community on these threats, Figure~\ref{fig:attack_trend} shows the cumulative number of \deleted{research studies published over time that specifically address attacks against voice authentication systems}\added{studies included in this survey by publication year, reflecting the growing research attention devoted to attacks against voice authentication systems}.

\begin{figure}[h]
    \centering
    \begin{tikzpicture}
    \begin{axis}[
        width=\linewidth, height=6.2cm,
        xlabel={Publication year}, ylabel={Cumulative studies},
        xmin=1989, xmax=2027, ymin=0, ymax=92,
        xtick={1990,1995,2000,2005,2010,2015,2020,2025},
        ytick={0,20,40,60,80},
        grid=major, grid style={gray!20},
        tick label style={font=\small}, label style={font=\small},
    ]
    \addplot[blue!70!black, thick, mark=*, mark size=1.1pt] coordinates {
        (1990,1) (1993,2) (1995,3) (1999,4) (2011,5) (2012,6) (2013,7) (2014,8)
        (2015,11) (2016,12) (2017,19) (2018,26) (2019,32) (2020,42) (2021,53)
        (2022,61) (2023,69) (2024,75) (2025,85) (2026,86)
    };
    \end{axis}
    \end{tikzpicture}
    \caption{\added{Cumulative number of studies included in this survey, by publication year.}}
    \label{fig:attack_trend}
\end{figure}

\textbf{Data poisoning attacks} inject malicious samples into the training pipeline to corrupt model behavior or embed hidden backdoor triggers. \textbf{Adversarial attacks} add imperceptible perturbations to input audio at inference time to induce misclassification. \textbf{Deepfake voice attacks} use speech synthesis or voice conversion to generate audio that mimics a target speaker's identity. \textbf{Adversarial spoofing attacks} combine synthesized audio with adversarial perturbations designed to simultaneously bypass both the speaker verifier and its anti-spoofing countermeasure \cite{almutairi2022review}.

\section{Attack Taxonomies and Threat Models}
\label{sec:taxonomies_and_threats}

This section provides a comprehensive framework for understanding and categorizing attacks against voice authentication systems. We first present a multi-dimensional taxonomy that compares attacks primarily along three axes---what the attacker knows, how the attack is delivered, and how broadly it generalizes across speakers and inputs---and further characterizes each attack by its objective, access level, composition complexity, and attack medium. We then describe the primary threat models that define different attack scenarios targeting VAS, establishing a foundation for the detailed technical discussions in subsequent sections.

\subsection{Attack Taxonomies}
\label{sec:taxonomies}

While prior taxonomies have focused on individual dimensions, notably attacker knowledge (white-box / grey-box / black-box)~\cite{lan2022adversarial} and attack objectives (targeted / untargeted)~\cite{xu2020adversarial}, they treat these axes independently without establishing how they compose across the full attack lifecycle. Our taxonomy extends these frameworks along two additional orthogonal dimensions: (1) \textbf{attack delivery channel} (over-the-line / over-the-air), which determines whether physical acoustic distortion must be modeled during attack design, and (2) \textbf{attack composition complexity} (Levels 0:4), which captures the sophistication of the attack generation pipeline from a single replay to adaptive feedback-driven optimization. Together, these dimensions enable cross-threat comparison, making explicit, for example, that a deepfake TTS attack is simultaneously a Level-1, black-box, targeted, over-the-line attack, something prior taxonomies cannot express within a unified framework.

\begin{figure*}[h]
    \centering
    \includegraphics[width=\textwidth]{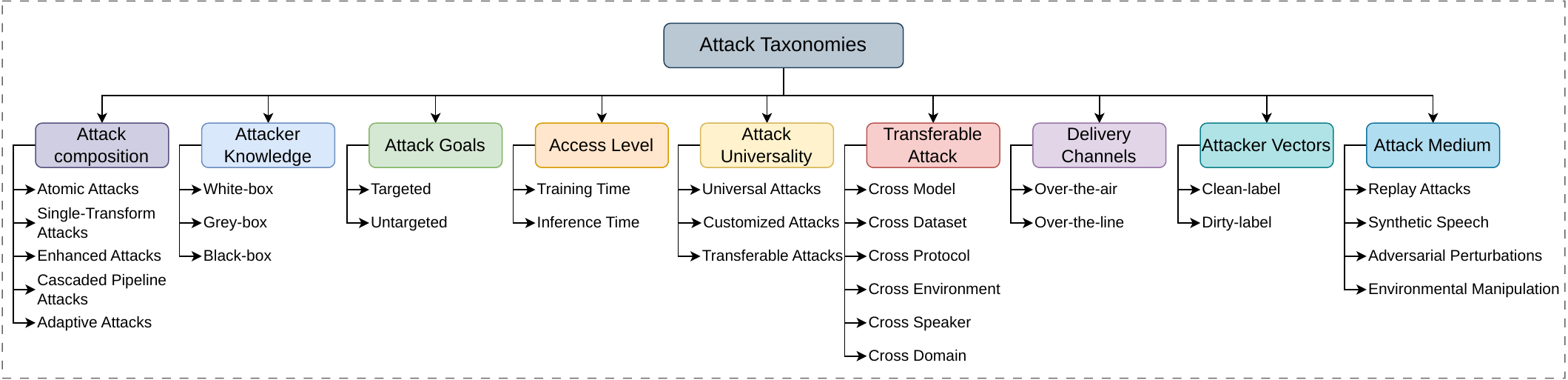}
    \caption{Comprehensive taxonomy of attacks on voice authentication systems, categorized by key factors including attacker knowledge, objectives, access level, perturbation type, transferability, medium, delivery channel, and attack vector.}
    \label{fig:taxonomies}
\end{figure*}

The landscape of attacks in voice authentication can be systematically categorized along several dimensions, as illustrated in Figure~\ref{fig:taxonomies}.

The \textbf{attack composition complexity} describes the sophistication of the attack generation pipeline through modular processing stages. At the simplest end, \textbf{atomic attacks (Level~0)} involve no computational transformation, for example replay of an unmodified recording or unaided human impersonation. \textbf{Single-transform attacks (Level~1)} apply one transformation method, such as standalone TTS synthesis, voice conversion without post-processing, or an isolated signal manipulation like pitch shifting. \textbf{Enhanced attacks (Level~2)} combine a primary transformation with obfuscation: TTS followed by reverberation, noise injection, or equalization; voice conversion with prosody modification; or synthesis augmented with adversarial perturbations. \textbf{Cascaded pipeline attacks (Level~3)} chain multiple transformations sequentially, for example TTS $\rightarrow$ voice conversion $\rightarrow$ post-processing, or synthesis $\rightarrow$ enhancement $\rightarrow$ anti-forensics obfuscation, to progressively refine quality. At the most sophisticated end, \textbf{adaptive attacks (Level~4)} incorporate feedback by adapting to system confidence scores, refining across multiple attempts, or using reinforcement learning to evolve their strategy through interaction with the target. This dimension is further refined by \textit{parallelism} (ensemble vs.\ sequential), \textit{component diversity} (homogeneous vs.\ heterogeneous pipelines), and \textit{feedback integration} (one-shot vs.\ iterative refinement).

The \textbf{attacker knowledge} dimension determines the information available to the adversary and shapes both attack strategy and complexity. In a \textbf{white-box} setting, the attacker has complete access to model architecture, parameters, gradients, and training data, enabling exact gradient computation and direct exploitation of architectural vulnerabilities; this scenario represents an upper bound on effectiveness and typically arises in academic research, insider threats, or attacks on publicly documented models. A \textbf{grey-box} attacker observes confidence scores or probability distributions but has no access to internal parameters or gradients, as is common with APIs that return ranked or scored responses; here, adversaries rely on score-based optimization such as zeroth-order methods or evolutionary algorithms, a realistic regime for deployed systems that expose richer feedback than binary decisions. The most restrictive setting is \textbf{black-box}, where the attacker sees only hard-label Accept/Reject responses and must resort to surrogate-model training, query-based boundary inference, or physical attacks such as replay and synthesis; this setting most closely matches production systems engineered to minimize information leakage.

The \textbf{attack goal} distinguishes attacks aimed at a specific identity from those aimed at generic failures. \textbf{Targeted attacks} seek to impersonate a particular legitimate user, which requires acquiring samples of that user and crafting an attack, through replay, voice conversion, TTS, or adversarial perturbation, that lands at the target's embedding location. This is the more demanding objective, since it requires navigating to a specific point in embedding space rather than merely crossing any decision boundary, and it poses the highest security risk by enabling unauthorized access to specific accounts. \textbf{Untargeted attacks}, in contrast, aim at system-level failures such as false acceptance, false rejection, misclassification, or confidence degradation, and only require crossing some decision boundary. Universal adversarial perturbations, noise injection, and entropy maximization fall in this category; although individually less severe, untargeted attacks at scale undermine system reliability and availability.

The \textbf{access level} determines which stage of the machine-learning pipeline the adversary can compromise. \textbf{Training-phase} attacks corrupt the model during development, typically through data poisoning, backdoor injection, malicious gradient updates in federated learning, or availability attacks that degrade overall accuracy. These vulnerabilities persist throughout deployment and are difficult to detect because poisoned models often behave normally on clean test data; such attacks generally require insider access or compromise of upstream data sources. \textbf{Inference-phase} attacks instead target the deployed model through adversarial perturbations, replay or synthesis spoofing, query-based probing, model evasion that exploits feature-extraction weaknesses, or physical manipulation of the acoustic environment. They require no access to training infrastructure and, partly for that reason, account for the majority of reported attacks against VAS.

\textbf{Attack universality} characterizes how broadly an attack generalizes across inputs, models, and scenarios. \textbf{Universal attacks} apply a single perturbation or methodology that compromises the system across a wide range of inputs, speakers, or models; universal adversarial perturbations (UAPs) cause misclassification when added to virtually any input and enable scalable attacks without per-sample tuning, with their very existence pointing to fundamental decision-boundary vulnerabilities. \textbf{Customized attacks}, by contrast, are tailored to a particular model, input, or target speaker; they typically require detailed knowledge of the target and achieve high success against that specific system but generalize poorly elsewhere, trading scalability for maximum per-target probability. \textbf{Transferable attacks} sit between these extremes: an attack crafted on a surrogate model remains effective against an unrelated target, exploiting shared decision boundaries or architectural patterns. Transferability is the principal mechanism by which black-box attacks succeed in practice, with transfer rates depending on architectural similarity, training-data overlap, shared features, and task alignment; ensemble-based optimization across multiple surrogates further improves transferability.

Within the broad notion of transferability, several orthogonal dimensions can be distinguished. \textbf{Cross-model} transfer occurs when an attack crafted on one architecture succeeds on others with different designs, for example, when an attacker uses an open-source surrogate to attack a proprietary system, and is strongest when models share input representations or training objectives. \textbf{Cross-dataset} transfer holds when perturbations crafted on one dataset remain effective on another despite distributional differences, suggesting that the exploited vulnerabilities reflect fundamental decision-boundary weaknesses rather than training-data artifacts; transfer is typically stronger between domain-similar datasets. \textbf{Cross-protocol} transfer applies across tasks, such as verification versus identification or closed-set versus open-set, and arises because these tasks share underlying embedding networks and differ chiefly in their decision layers. \textbf{Cross-environment} transfer measures whether adversarial audio survives noise, channel shifts, or hardware variation, which is critical for over-the-air attacks subject to physical playback and re-recording; robustness here is typically obtained through expectation-over-transformation (EOT), room-impulse-response simulation, and noise injection during optimization. \textbf{Cross-speaker} transfer captures whether a single methodology spoofs multiple enrolled users beyond the original target, indicating systemic rather than individual vulnerabilities. Finally, \textbf{cross-domain} transfer asks whether attacks generalize between synthetic and natural speech or between distinct speech-generation methods (TTS, voice conversion, human speech), an increasingly important property as generative speech technologies proliferate.

The \textbf{attack delivery channel} specifies the physical or digital pathway by which malicious audio reaches the system. \textbf{Over-the-line} attacks deliver inputs digitally, through file uploads, VoIP calls, APIs, or compromised communication links, and operate entirely in the digital domain via direct API submission, telephony transmission, protocol exploitation, or man-in-the-middle interception. This channel offers precise signal control and easy scaling but also exposes the attack to digital forensics, cryptographic authentication, and traffic-anomaly detection. \textbf{Over-the-air} attacks instead play adversarial audio aloud, relying on a microphone to capture it; this category includes replay attacks, broadcast of TTS or VC output, ultrasonic and inaudible attacks, and acoustic adversarial examples engineered to survive physical propagation. Such attacks must contend with acoustic distortion, ambient noise, and hardware variation, which is why recent work increasingly incorporates room-impulse-response simulation during optimization.

Within data poisoning specifically, the \textbf{attacker vector} describes how the adversary manipulates the training pipeline. \textbf{Dirty-label poisoning} modifies both the input $x_i$ and its label $y_i$, for example by labeling speaker A as speaker B, marking spoofed samples as genuine, or planting targeted backdoors. Although this approach is highly effective, the inconsistency between content and label often makes it detectable through data auditing and consistency checks. \textbf{Clean-label poisoning} is significantly stealthier: only the input $x_i$ is subtly altered while the label remains correct, so the data passes label inspection and standard hygiene checks. Detection in this regime typically requires more sophisticated tools such as influence-function analysis or anomaly detection in embedding space. The effectiveness of both variants depends on the poisoning ratio and on the model's intrinsic robustness to label noise.

Finally, the \textbf{attack medium} characterizes the fundamental nature of the malicious audio signal. \textbf{Replay attacks} re-broadcast pre-recorded speech of the target, directly, partially, or after light editing, and remain the simplest spoofing method, requiring only target speech and a playback device; they are typically detectable through playback-recording artifacts, missing natural variability, or mismatched room acoustics. \textbf{Synthetic speech} is generated by AI-based TTS or VC pipelines that imitate a target's voice, with modern neural vocoders and few-shot adaptation dramatically raising fidelity; countermeasures include synthetic-speech detection and liveness verification. \textbf{Adversarial perturbations} are carefully optimized additive noise designed to cause targeted misclassification while remaining imperceptible, generated through gradient-based methods such as FGSM, PGD, and C\&W, or under psychoacoustic constraints, with imperceptibility, physical robustness, and transferability remaining the principal open challenges; common defenses include adversarial training, input transformation, and certified robustness. \textbf{Environmental manipulation} attacks exploit the physical layer itself, hiding adversarial signals in ambient noise, abusing microphone non-linearities through ultrasonic or ``dolphin'' attacks, or jamming the channel, and are countered with robust audio front-ends (e.g.,\ ultrasonic filtering) and multimodal sensing.

Table~\ref{tab:taxonomy_positioning} synthesizes how each of the four threat vectors analyzed in this survey positions along the six taxonomy dimensions defined above, providing a single-page reference that supports systematic cross-threat comparison. For instance, the table makes explicit that a deepfake TTS attack is simultaneously black-box, targeted, inference-phase, and over-the-line, a multi-dimensional characterization that single-axis taxonomies cannot express.

\begin{table*}[h]
\centering
\caption{Positioning of the four attack categories along the proposed taxonomy dimensions.}
\label{tab:taxonomy_positioning}
\resizebox{\textwidth}{!}{%
    \begin{tabular}{lllllll}
    \toprule
    \textbf{Attack} & \textbf{Knowledge} & \textbf{Goal} & \textbf{Phase} & \textbf{Channel} & \textbf{Complexity} & \textbf{Universality} \\
    \midrule
    Data Poisoning & White/Grey & Targeted/Untargeted & Training & Over-the-line & L1--L2 & Customized \\
    Adversarial & White/Grey/Black & Targeted/Untargeted & Inference & Both & L2--L4 & Universal/Transferable \\
    Deepfake & Black & Targeted & Inference & Both & L1--L3 & Transferable \\
    Adv.\ Spoofing & White/Grey & Targeted\added{/Untargeted} & Inference & Both & L3--L4 & Customized \\
    \bottomrule
    \end{tabular}%
}
\end{table*}

\subsection{Threat Models in Voice Authentication Attacks}
\begin{figure*}[h]
    \centering
    \includegraphics[width=\textwidth]{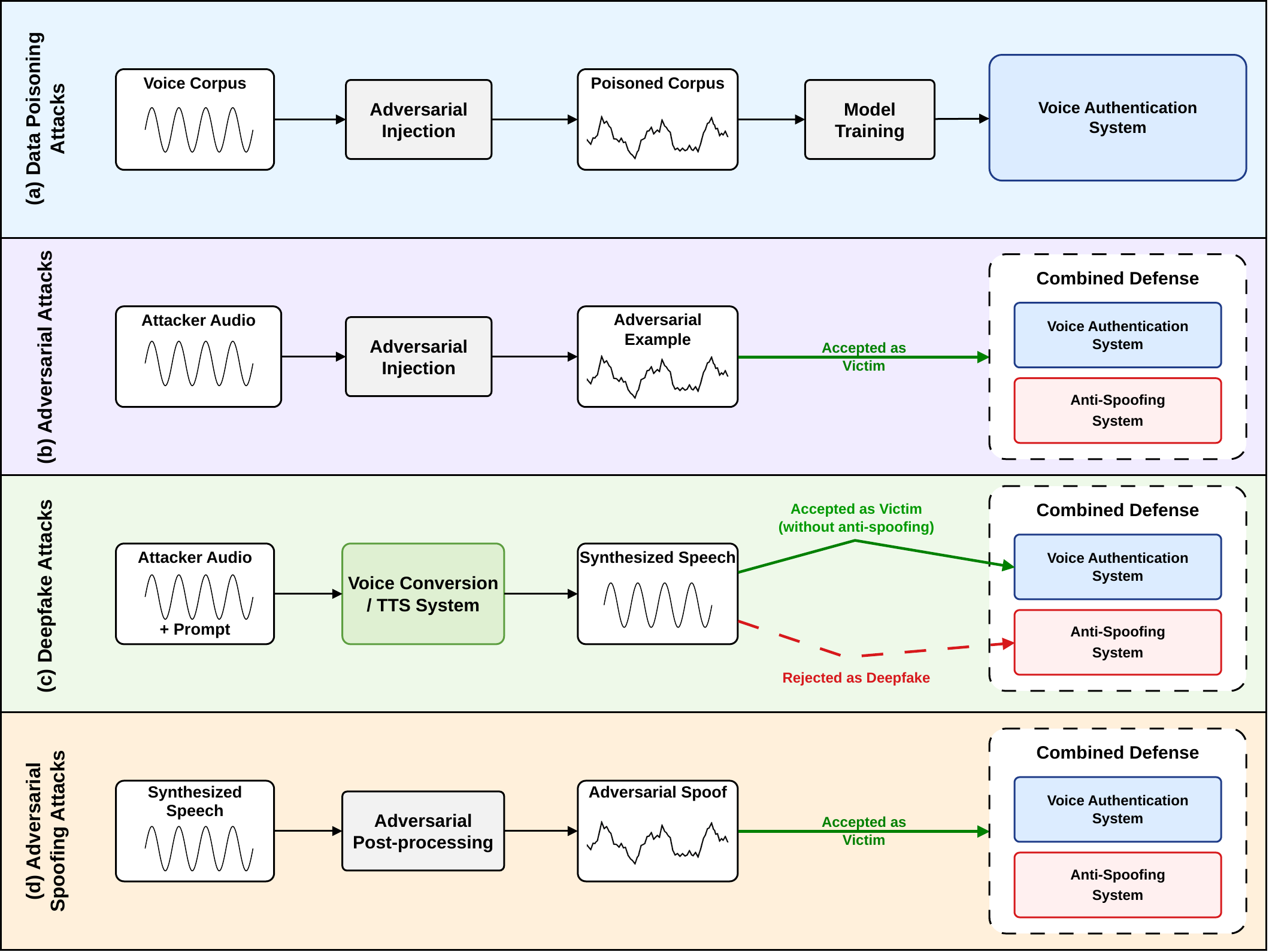}
    \caption{Attacks Threat Models (a) Data Poisoning Attacks. (b) Adversarial Attacks. (c) Deepfake Audio Attacks. (d) Adversarial Spoofing Attacks.}
    \label{fig:attacks}
\end{figure*}

Voice authentication systems are vulnerable to a variety of sophisticated threats, each exploiting different stages of the machine learning lifecycle or system architecture. Figure~\ref{fig:attacks} illustrates the relationship between these attack types.

\textbf{Data poisoning attacks} target the training phase of VAS by injecting malicious or manipulated data into the model's training pipeline to corrupt its learned representations or behavior.

As discussed in Section~\ref{sec:taxonomies}, the attacker's knowledge can vary: in a \textit{white-box} setting, they may craft highly optimized poisoned inputs using full model access; in a \textit{grey-box} scenario, poisoning is guided by partial system knowledge; and in a \textit{black-box} setting, attacks rely on limited feedback or surrogate datasets. Poisoning goals may be \textit{targeted}, such as embedding a backdoor to impersonate a specific user, or \textit{untargeted}, aiming to degrade the model's overall performance. Attack vectors include both \textit{clean-label} and \textit{dirty-label} strategies, depending on whether the injected samples are correctly or incorrectly labeled. Further technical details and relevant literature will be discussed in Section~\ref{sec:data_poisoning}.

\textbf{Adversarial attacks} exploit vulnerabilities during the inference phase by introducing carefully crafted perturbations to input audio. These perturbations are typically imperceptible to humans and constrained under $L_p$-norm bounds (e.g., $L_2$, $L_\infty$), allowing them to bypass detection while manipulating model predictions.

According to the taxonomy in Section~\ref{sec:taxonomies}, adversarial attacks may occur under \textit{white-box} conditions, where the attacker leverages full model access to compute gradients; \textit{grey-box} settings, with partial system knowledge; or \textit{black-box} scenarios, relying on queries or transferability. The objective may be \textit{targeted}, forcing the system to accept an adversary as a specific user, or \textit{untargeted}, simply inducing misclassification or denial of access. Effective adversarial examples maintain a trade-off between attack success and perceptual naturalness. Further technical details and relevant literature will be discussed in Section~\ref{sec:adv_attacks}.

\textbf{Deepfake voice attacks} exploit generative models to synthesize speech that mimics a target speaker, aiming to deceive VAS. The success of these attacks depends largely on the attacker's available resources. 

The attacker's goal is typically \textit{targeted}, seeking to impersonate a specific speaker by generating audio that meets or exceeds the similarity threshold $\tau$ required for authentication. Common techniques include \textit{VC}, which transforms one speaker's voice to sound like another's, and \textit{TTS}, where new phrases are generated using a cloned voice profile. These attacks challenge a system's ability to distinguish between genuine and synthetic input, particularly as deep generative models continue to improve. Further technical details and relevant literature will be discussed in Section~\ref{sec:deepfake_attacks}.

\textbf{Adversarial spoofing attacks} emerge when adversarial perturbations are designed to deceive both VAS and anti-spoofing CMs. These attacks typically utilize the output of TTS or VC systems, which are already capable of targeting VAS, and further adapt them to bypass anti-spoofing CMs as well.

These attacks can follow a \textit{white-box} paradigm, where the adversary has access to gradients of both the speaker recognition model $f_\theta$ and the spoof detector $d_\phi$, allowing for coordinated optimization across both targets. These dual-purpose attacks typically involve minimizing a joint loss function: $\mathcal{L}_{\text{auth}} + \lambda\mathcal{L}_{\text{spoof}}$, where $\lambda$ adjusts the emphasis between breaking authentication and bypassing spoof detection. In Grey-box settings, attackers utilize query-based methods on confidence scores. In strict Black-box settings, attackers must rely entirely on surrogate model transferability or universal adversarial filters. The attack goal may be either \textit{targeted} or \textit{untargeted}, depending on whether the adversary seeks impersonation or generalized evasion. This multi-objective threat landscape underscores the need for robust, multi-layered defense strategies. Further technical details and relevant literature will be discussed in Section~\ref{sec:adv_attacks_cms}.

\section{Data Poisoning Attacks}
\label{sec:data_poisoning}

Data poisoning attacks are a critical security concern for VAS, where adversaries manipulate training data to compromise a model's integrity or availability \cite{jagielski2018manipulating, biggio2012poisoning, ling2019deepsec}. With the increasing reliance on third-party data sources, including cloud-based services and web scraping, the risk of unintentionally incorporating poisoned data into training pipelines has increased significantly, particularly in the audio domain, where model training is computationally expensive and data-hungry \cite{ling2019deepsec, jagielski2018manipulating}. A summary of these attacks is illustrated in Table~\ref{tab:data_poisoning_attacks}.

\subsection{Problem Formulation}
We formally define data poisoning against a voice authentication system as a bi-level optimization problem. Let $\mathcal{D}_c = \{(x_i, y_i)\}$ represent the clean training dataset and $\mathcal{D}_p = \{(x_j, y_j)\}$ represent the injected poisoned dataset. The model $f_\theta$ is trained on the combined dataset $\mathcal{D}_{train} = \mathcal{D}_c \cup \mathcal{D}_p$.
The attacker's objective is to construct $\mathcal{D}_p$ such that the trained model parameters $\theta^*$ minimize an adversarial objective $\mathcal{A}$ (e.g., maximizing classification error on a specific target or the entire validation set $\mathcal{D}_{val}$), subject to a poisoning budget constraint $\epsilon$ that limits the infection rate.
\begin{equation}
    \begin{aligned}
    \max_{\mathcal{D}_p} \quad & \mathcal{A}(\theta^*, \mathcal{D}_{val}) \\
    \textrm{s.t.} \quad & \theta^* \in \arg\min_\theta \sum_{(x,y) \in \mathcal{D}_c \cup \mathcal{D}_p} \mathcal{L}_{train}(f_\theta(x), y), \\
    & |\mathcal{D}_p| \leq \epsilon |\mathcal{D}_c|
    \end{aligned}
\end{equation}
where $\mathcal{L}_{train}$ is the standard training loss function (e.g., Cross-Entropy) and $\epsilon$ is the maximum allowable poisoning ratio.
\subsubsection*{Feature Collision (Clean-Label) Model}
For Feature Collision (Clean-label) attacks, the attacker optimizes a specific poisoned sample $x_p$ starting from a base instance $x_b$. The goal is to minimize the distance between the poison sample's feature representation $g(x_p)$ and a target victim's representation $g(x_t)$, while constraining the input to resemble the base instance $x_b$:
\begin{equation}
x_p = \arg\min_x \; \lVert g(x) - g(x_t) \rVert_2^2 + \lambda \lVert x - x_b \rVert_2^2
\end{equation}
where $g(\cdot)$ denotes the feature extraction layer of the model, and $\lambda$ is a regularization coefficient balancing feature collision success and perceptual similarity.

\subsection{Untargeted Data Poisoning Attacks}
Untargeted data poisoning attacks seek to degrade the overall performance of a VAS without directing the model toward any specific misclassification outcome. In the audio domain, these attacks typically exploit the model’s reliance on particular feature representations or training data statistics. The following studies illustrate the range of techniques and share a common limitation: most require unrealistically large poisoning budgets or white-box access that is difficult to obtain in production settings.

Fowl et al. \cite{fowl2021adversarial} demonstrated this by injecting adversarial examples into training sets. Their method uses crafted Projected Gradient Descent (PGD) perturbations on a substitute CNN to generate semantically valid but misleading inputs. Although this was effective in lab settings, it requires access to the pretrained model and full-batch training data, which often breaks down in streaming or privacy-preserving voice data pipelines, limiting real-world applicability.

Huang et al. \cite{huang2021unlearnable} proposed an alternative approach using unlearnable examples, which suppress informative features in training data, disrupting representation learning in audio models. Although this clean-label method cleverly avoids detection through label inspection, its effectiveness heavily depends on the model's inability to recover signal through preprocessing or data augmentation, standard practices in modern voice pipelines. This dependency makes the attack less transferable.

WaveFuzz by Ge et al. \cite{ge2022wavefuzz} takes clean-label poisoning a step further by changing Mel-Frequency Cepstral Coefficients (MFCC) features to mislead speech models without affecting human perception. Their method lowers accuracy in end-to-end training, but it relies on certain feature extraction techniques, like MFCC. It also has less effect on stronger architectures, such as ResNet-18 and Kaldi \cite{Povey2011Kaldi}, which shows its limited ability to generalize. Additionally, the need to poison at least 10\% of the training data brings up concerns about practicality in real-world use.

A more targeted attack is SyntheticPop \cite{jamdar2025syntheticpop}, which exploits liveness detection systems that use pop noise, like VoicePop \cite{wang2019voicepop}, by embedding low-frequency sine-wave ``pops'' into training samples. This manipulation at the feature level causes Gammatone Frequency Cepstral Coefficients (GFCC)-based detectors to fail. The accuracy drops from 69\% to 14\% with only 20\% poisoning. While this approach is impressive, it assumes white-box access to the target’s audio pipeline. It does not work against systems that use different feature sets or basic denoising, which lowers its threat level in more diverse or well-defended environments.

Across these works, a pattern emerges. Untargeted poisoning methods for audio depend heavily on certain model assumptions, such as known architectures, fixed features, and large-scale poisoning. This heavy dependence weakens their transferability and realism. The field needs more flexible and discreet poisoning strategies that consider audio-specific issues like preprocessing pipelines, limited labels, and data streaming.

\subsection{Targeted Data Poisoning Attacks}
Targeted data poisoning attacks aim to manipulate machine learning models to misclassify specific targets at inference time, without degrading overall performance. These attacks pose a critical risk to biometric systems, where identity spoofing can have serious consequences.

One influential method is the feature collision poisoning introduced by Shafahi et al. \cite{shafahi2018poison}. In this approach, adversarial examples are created to resemble target samples in the feature space while still looking harmless in the input space. This technique achieved a 100\% success rate in transfer learning scenarios using a single poisoned sample, with a negligible accuracy loss. However, its scalability to end-to-end models is limited; achieving comparable success required up to 50 poison instances, enhanced using a 30\% opacity watermark of the target, highlighting a trade-off between attack stealth and practicality.

Expanding into the voice domain, Li et al. \cite{li2023defend} showed a two-phase attack on speaker recognition models like DeepSpeaker \cite{deepspeaker}. They used label flipping during training and later mixed attacker samples with victim identities in the enrollment process. This method achieved up to 90\% success while keeping overall accuracy high. However, their threat model assumes access to the training pipeline and uses a large number of enrollment samples (10). This does not reflect real-world practices, where systems usually use fewer samples (1–3) \cite{amazonvoiceid, googlespeakerid}. Additionally, the attack's effectiveness depends on ideal conditions, which may not be present in production systems. This raises doubts about its practical use.

Addressing these limitations, Mohammadi et al. \cite{mohammadi2024securing} proposed a more practical evaluation of poisoning threats. They focused on stealthy, small-scale poisoning with 5\% data contamination. They introduced a hybrid CNN and k-Nearest Neighbors (k-NN) defense that exceeded earlier defenses (The Guardian \cite{li2023defend}) in detection accuracy on real-world datasets, such as VoxCeleb \cite{nagrani2020voxceleb}. Their approach balances robustness and usability, but it still assumes that the attacker can inject poisoned samples labeled as the victim. This assumption may not match typical production systems where training happens offline and is separated from user-specific data.

The targeted variants tell a similar story from the other direction: they achieve high success against specific identities, but only under conditions---full access to the training process and knowledge of the model internals---that rarely hold in practice. Bridging this gap will require threat models built around realistic deployment constraints rather than controlled-laboratory assumptions.

\subsection{Backdoor Attacks}
\textbf{Backdoor attacks} in machine learning involve inserting harmful behaviors into a model during training. These behaviors activate only when specific, often subtle, triggers are present. Unlike data poisoning, backdoor attacks aim to maintain high accuracy on clean inputs while secretly allowing targeted misclassification. This threat is especially pronounced in speaker verification and identification systems, where backdoors can silently override identity checks.

Early work by Liu et al. \cite{liu2018trojaning} introduced a foundational Trojaning method, embedding triggers by manipulating high-activation neurons and reverse-engineering synthetic training data. This attack achieved near-perfect success across domains (including face recognition and driving systems) with minimal computational overhead and no loss in model accuracy. However, it relied on full white-box access, which limits applicability in production environments.

Building on this, Xu et al. \cite{xu2021detecting} proposed a broader taxonomy of backdoor strategies—modification, blending, and parameter/latent attacks—each targeting different modalities. While successful in vision and speech domains, attacks like blending were less effective due to feature interference, and parameter-based methods struggled with non-vision tasks. Crucially, these attacks required access to the training dataset, reducing their stealth in practical deployment settings.

As attacks advanced, researchers explored novel, audio-specific triggers that exploit unique acoustic properties. For instance, Zhai et al. \cite{zhai2021backdoor} designed a clustering-based approach for speaker verification, using low-energy spectral noise per cluster. Their attack retained stealth (minimal Equal Error Rate (EER) elevation) while outperforming traditional backdoors like BadNets \cite{gu2019badnets}. However, it traded off simplicity for operational complexity, requiring multiple trigger variants and increased inference-time burden.

A more subtle and innovative approach was introduced by Schoof et al. \cite{schoof2024emoback} in EmoBack, which utilized emotional prosody (e.g., pitch and speech rate) as a dynamic trigger. This strategy achieved an Attack Success Rate (ASR) up to 98.9\% while preserving clean accuracy. Yet, its performance varied by emotion and model architecture, and it required higher poisoning rates (5–10\%), reducing stealth. Defensive pruning degraded the attack partially, underscoring the ongoing cat-and-mouse nature of mitigation.

Pushing stealth to the extreme, Ye et al. \cite{ye2023fake} leveraged voice conversion using FreeVC \cite{li2023freevc} to poison training data with speaker-morphed utterances, achieving 99.94\% ASR with only 1\% poisoning. These samples preserved the original semantics while replacing the speaker identity, a significant step toward imperceptible attacks. However, their generalizability was constrained: effectiveness sharply dropped when voiceprints did not match across train and test speakers, and success remained architecture-dependent (e.g., RawNet3\cite{rawnet3} \textless X-vectors \cite{xvectors}).

In parallel, Cai et al. \cite{cai2022pbsm} introduced inaudible triggers via ultrasonic frequencies (18–22 kHz) and pitch modifications (PBSM), exploiting psychoacoustic masking. These triggers were effective (ASR up to 100\%) yet vulnerable to environmental noise and hardware variability. Follow-up analysis by Mohammadi et al. \cite{mohammadi2025mitigating} demonstrated that these spectral attacks can bypass traditional defenses but fail under common preprocessing techniques like downsampling or non-linear transformations.

This growing class of non-semantic triggers, from environmental sounds (e.g., keyboard clicks) to emotional embeddings, offers high success rates but raises concerns about practical deployment. Triggers must survive real-world transformations and remain undetectable across devices, which is not always feasible. Emotional triggers, for example, may be misclassified due to natural variation, leading to false positives.

\begin{table*}[h]
    \centering
    \caption{Summary of Data Poisoning and Backdoor Attacks on Speaker Recognition Systems}
    \label{tab:data_poisoning_attacks}
    \resizebox{\textwidth}{!}{%
    \begin{tabular}{p{1cm}p{0.8cm}p{2.8cm}p{2.8cm}p{2cm}p{1.3cm}p{5cm}}
        \toprule
        \textbf{Study} & \textbf{Year} & \textbf{Mechanism} & \textbf{Target} & \textbf{Poisoning Rate} & \textbf{ASR} & \textbf{Notes / Limitations} \\
        \midrule
        \multicolumn{7}{l}{\textbf{Untargeted Data Poisoning}} \\
        \midrule
        \cite{fowl2021adversarial} & 2021 & Adv. Examples & CNN-SV & — & — & Requires pre-trained model and full dataset access \\
        \addlinespace
        \cite{huang2021unlearnable} & 2021 & Unlearnable Examples & Audio CNNs & — & — & Vulnerable to preprocessing/augmentation \\
        \addlinespace
        \cite{ge2022wavefuzz} & 2022 & Clean-label flip & MFCC-based models & $\geq 10\%$ & — & Less effective on ResNet/FFT, Kaldi shows robustness \\
        \addlinespace
        \cite{jamdar2025syntheticpop} & 2025 & Synthetic Pop Noise & GFCC + SVM & 20\% & 95\% & White-box, pop-noise specific, mitigated by suppression \\
        \addlinespace
        \midrule
        \multicolumn{7}{l}{\textbf{Targeted Data Poisoning}} \\
        \midrule
        \cite{shafahi2018poison} & 2018 & Feature Collision & MFCC-based models & — & — & Classical attack, less explored in audio domain \\
        \addlinespace
        \cite{li2023defend} & 2023 & Enrollment Injection & DeepSpeaker & 10\% Train. - 50\% Enroll & 90\% & Unrealistic assumptions about enrollment data access \\
        \addlinespace
        \cite{mohammadi2024securing} & 2024 & Low-rate Label Flip & LibriSpeech, VoxCeleb & 5\% & — & Assumes attacker controls training, limited in deployed systems \\
        \addlinespace
        \midrule
        \multicolumn{7}{l}{\textbf{Backdoor Attacks}} \\
        \midrule
        \cite{liu2018trojaning} & 2018 & Neuron Trigger & Neural Networks & — & 95.5--100\% & Full white-box access required \\
        \addlinespace
        \cite{xu2021detecting} & 2021 & Blending, Parameter & Speaker Recognition Models & — & 98.66\% & Detectable triggers, poor transfer to text/tabular data \\
        \addlinespace
        \cite{zhai2021backdoor} & 2021 & Clustering Trigger & SV Systems & 15\% & 63.5\% & Limited stealth at high rates, cluster sensitivity \\
        \addlinespace
        \cite{cai2022pbsm} & 2022 & Ultrasonic Frequency Trigger & Frequency-sensitive Models & \textless5\% & 95--100\% & Hardware sensitivity, fails under downsampling \\
        \addlinespace
        \cite{ye2023fake} & 2023 & Voice Conversion Trigger & VoxCeleb1-based Models & 1\% & 99.94\% & Fails for mismatched voices, task-dependent performance \\
        \addlinespace
        \cite{schoof2024emoback} & 2024 & Emotional Prosody Trigger & ResNet, X-vector, ECAPA & 5–10\% & 98.9\% & Emotion-dependent, pruning reduces efficacy \\
        \addlinespace
        \cite{mohammadi2025mitigating} & 2025 & Backdoor Defense & Spectrogram-based Detectors & N/A & — & Reveals PBSM stealth limits, preprocessing counters attack \\
        \bottomrule
    \end{tabular}%
    }
\end{table*}

\section{Adversarial Attacks}
\label{sec:adv_attacks}

Adversarial attacks introduce subtle, often imperceptible, perturbations into speech signals to mislead Speaker Recognition Systems (SRS). Unlike spoofing attacks, which typically involve synthesized or replayed speech, adversarial attacks directly manipulate the input data to exploit and alter the model's decision boundaries. As speaker recognition technology has transitioned from traditional methods, such as Gaussian Mixture Model-Universal Background Model (GMM-UBM) pipelines, to deep learning-based architectures like x-vectors \cite{xvectors} and ECAPA-TDNN \cite{desplanques2020ecapa}, the sophistication of attack strategies has advanced in parallel, evolving from purely digital manipulations to practical over-the-air implementations. Table~\ref{tab:adversarial_attacks} summarizes key adversarial attacks on speaker recognition systems.

\subsection{Problem Formulation}
Let $f_\theta: \mathcal{X} \rightarrow \mathcal{Y}$ be a speaker recognition model mapping audio $x \in \mathbb{R}^T$ to a label $y$. The goal of an adversarial attack is to find a perturbation $\delta$ that induces misclassification, subject to a constraint bounded by $\epsilon$ under an $L_p$-norm.

\begin{equation}
    \begin{aligned}
        \min_{\delta} \quad & \lVert \delta \rVert_p \\
        \textrm{s.t.} \quad & f_\theta(x + \delta) \neq y_{true} \quad (\text{Untargeted}) \\
        \text{or} \quad & f_\theta(x + \delta) = y_{target} \quad (\text{Targeted}) \\
        & x + \delta \in [-1, 1]^T \quad (\text{Valid Audio Range})
    \end{aligned}
    \end{equation}
    \subsubsection*{Attack Algorithms}
    To solve this constrained optimization, gradient-based methods are commonly employed:
    \textbf{Fast Gradient Sign Method (FGSM):} A single-step approximation using the sign of the gradient of the loss function $\mathcal{L}$:
    \begin{equation}
    x_{adv} = x + \epsilon \cdot \text{sign}(\nabla_x \mathcal{L}(f_\theta(x), y))
    \end{equation}
    \textbf{Projected Gradient Descent (PGD):} An iterative formulation that applies FGSM multiple times with a smaller step size $\alpha$, projecting the result back into the $\epsilon$-ball $\mathcal{S}$ at each step $t$:
    \begin{equation}
    x_{t+1} = \Pi_{\mathcal{S}} \left( x_t + \alpha \cdot \text{sign}(\nabla_{x} \mathcal{L}(f_\theta(x_t), y)) \right)
\end{equation}
where $\Pi_{\mathcal{S}}$ is the projection operator ensuring $\lVert x_{t+1} - x \rVert_p \leq \epsilon$.

\subsection{Foundational Work}
The concept of adversarial audio in speaker verification was first explored by Hautamäki et al. \cite{hautamaki13_interspeech}, who employed professional impersonators to target Speaker Verification (SV) systems. Although the false acceptance rate (FAR) increased only slightly from 9.03\% to 11.61\%, these mimicry attacks demonstrated that human impersonation poses a less significant threat than computational, synthetic methods.

A pivotal shift occurred with the work of Kreuk et al. \cite{kreuk2018fooling}, who applied the Fast Gradient Sign Method (FGSM) to generate targeted adversarial perturbations for end-to-end neural models. This attack reduced accuracy from 85.5\% to 37.5\% on the YOHO dataset \cite{campbell1995testing} and increased the false positive rate from 4.88\% to 94.63\%. Furthermore, the adversarial examples demonstrated transferability across different datasets and model architectures, highlighting a systemic vulnerability in neural SRS.

\subsection{White-Box Attacks}
White-box attacks represent the most powerful threat model, granting the adversary full access to model parameters and gradients. While this assumption limits real-world applicability, production systems rarely expose their internals, white-box results establish an upper bound on attack effectiveness and reveal fundamental vulnerabilities in the model's decision geometry. The studies below trace the evolution of these attacks from feature-space manipulations to physically realizable perturbations.

Li et al.~\cite{li2020adversarial} first extended these gradient-based attacks to traditional systems, including GMM-UBM and i-vector models. By applying FGSM to the feature space, they increased the FAR to nearly 100\% using imperceptible perturbations. Building on this, Li et al.~\cite{li2020practical} designed practical attacks against the state-of-the-art x-vector system, with both untargeted and targeted objectives. A key innovation was the integration of an estimated Room Impulse Response (RIR) to craft perturbations robust enough to withstand real-world playback distortions. The approach achieved a 98\% attack success rate (ASR) in digital simulations and 50\% in over-the-air physical playback tests, marking a significant advance in physically realizable adversarial attacks.

Further exploring these vulnerabilities, Joshi et al.~\cite{joshi2021study} analyzed modern x-vector-based systems under white-box conditions. They demonstrated that iterative attacks such as the Basic Iterative Method (BIM), Projected Gradient Descent (PGD), and Carlini-Wagner (CW) could reduce classification accuracy to nearly zero, even with a minimal perturbation budget ($\ell_\infty \le 0.001$). In parallel, Zhang et al.~\cite{zhang2020voiceprint} developed VMask, which utilizes psychoacoustic masking to craft perturbations that are imperceptible to humans but effective against both commercial and open-source systems. However, its real-world applicability was hindered by sensitivity to playback hardware and environmental noise. These findings collectively underscore the fragility of deep speaker representations against a variety of white-box threats.

Jati et al. \cite{jati2021adversarial} conducted a comprehensive study on the vulnerability of deep speaker recognition systems to adversarial attacks, evaluating standard algorithms like Fast Gradient Sign Method (FGSM), Projected Gradient Descent (PGD), and Carlini-Wagner (C\&W) in white-box settings. Their work establishes a critical baseline by demonstrating that while C\&W attacks generate the most imperceptible perturbations (highest Signal-to-Noise Ratio), PGD-based adversarial training proves to be the most effective defense, albeit with a trade-off in clean data performance. However, the study is primarily limited to digital, white-box scenarios and does not extensively evaluate robustness under real-world over-the-air conditions or black-box constraints.

Sankala et al. \cite{sankala2025adversarial} proposed a universal adversarial perturbation method specifically targeting Text-Dependent Speaker Verification (TD-SV) systems, challenging the assumption that fixed passphrases provide inherent security. Their approach generates a single, utterance-agnostic perturbation that allows an attacker to bypass verification using an arbitrary wrong phrase, removing the need for the attacker to have access to the victim's enrollment audio. While this highlights a severe vulnerability in password-based systems, the attack's effectiveness was primarily evaluated on specific TD-SV architectures, and its transferability to large-vocabulary or text-independent systems remains limited.

\subsection{Black-Box Attacks}
The FakeBob attack~\cite{chen2021real} demonstrated a practical black-box approach by combining the Natural Evolution Strategy (NES) and BIM to achieve a targeted ASR exceeding 90\% against i-vector systems. However, its practicality was limited by its high query cost, requiring thousands of model queries per sample.

To address the high query requirement, Chen et al.~\cite{chen2023qfa2sr} proposed QFA2SR, a query-free black-box attack. It leverages ensemble-based surrogate models and knowledge distillation to generate adversarial examples without direct access to the target system. This method achieved an 89.3\% ASR against commercial APIs, including Microsoft Azure, but its reliance on complex surrogate model training requires significant computational resources.

\subsection{Physical Over-the-Air Attacks}
To bridge the gap between digital simulations and real-world applications, Chen et al.~\cite{chen2022as2t} proposed AS2T, a generalizable transfer-based framework targeting SV, CSI, and OSI. By utilizing multi-task learning and RIR simulation, AS2T achieved an ASR of up to 98.5\% in white-box settings and demonstrated strong cross-task generalization.

PhyTalker~\cite{chen2022push} extended adversarial attacks to the real-time physical domain by injecting precisely timed, phoneme-level perturbations during live playback. PhyTalker achieved an 85\% targeted ASR in over-the-air tests without requiring access to the target model's parameters, demonstrating that physical attacks can compromise robust systems in real-time.

Ahmed et al. \cite{ahmed2023tubes} introduced a novel analog adversarial attack that manipulates the acoustic resonance of a speaker's voice using physically tuned PVC pipes to impersonate a target. By employing a greedy optimization algorithm to determine the precise tube dimensions required to shift formants, this method achieved a 60\% attack success rate against text-independent systems without relying on any electronic synthesis or digital manipulation. A major advantage of this approach is its ability to bypass digital spoofing detectors that look for synthetic artifacts; however, its practicality is constrained by the need for physical devices and the requirement that the attacker's natural pitch is close to the target's.

Xiao et al. \cite{xiao2025timbreadv} developed ``TimbreAdv," a framework for generating asynchronous, universal adversarial perturbations that target the timbre features of a speaker's voice in real-time. This method creates a single perturbation capable of deceiving systems across different speakers and texts, achieving a 95.1\% success rate in digital settings and maintaining robustness after over-the-air playback. Despite its high transferability and real-time applicability, the attack relies on the assumption that timbre features are the primary discriminators, potentially limiting its success against systems that heavily weigh prosodic or linguistic content.

\begin{table*}[htbp]
\centering
\caption{Comparison of Key Adversarial Attacks on Speaker Recognition Systems.}
\label{tab:adversarial_attacks}
\renewcommand{\arraystretch}{1.3}
\resizebox{\textwidth}{!}{
\begin{tabular}{p{2cm}p{0.7cm}p{2cm}p{2.3cm}p{2.5cm}p{3cm}p{5cm}}
    \toprule
    \textbf{Reference} & \textbf{Year} & \textbf{Attack Type} & \textbf{Target System} & \textbf{Method} & \textbf{Key Results} & \textbf{Key Findings / Limitations} \\
    \midrule
    Hautamäki et al. \cite{hautamaki13_interspeech} & 2013 & Human Mimicry & GMM-UBM, i-vector & Professional impersonation & FAR: 9.03\% $\rightarrow$ 11.61\% & Less effective than computational synthetic methods. \\
    \midrule
    \multicolumn{7}{l}{\textbf{White-Box Attacks}} \\
    \midrule
    Kreuk et al. \cite{kreuk2018fooling} & 2018 & Gradient-based, Transferable & End-to-end Neural Nets & FGSM on MFCCs & Accuracy: 85.5\% $\rightarrow$ 37.5\% \newline FPR: 4.88\% $\rightarrow$ 94.63\% & Demonstrated high transferability across models and datasets. \\
    \addlinespace
    Li et al. \cite{li2020adversarial} & 2020 & Gradient-based, Transferable & GMM-UBM, i-vector & FGSM on feature space & FAR $\approx$ 100\% (white-box) \newline EER up to 50\% (black-box) & Achieved high success with imperceptible noise; transferable. \\
    \addlinespace
    Li et al. \cite{li2020practical} & 2020 & Over-the-Air (OTA) & x-vector & Gradient optimization + RIR estimation & ASR: 98\% (digital), 50\% (OTA) & Focused on practical, physically robust perturbations. \\
    \addlinespace
    Joshi et al. \cite{joshi2021study} & 2021 & Iterative Gradient-based & x-vector & BIM, PGD, CW attacks & Accuracy $\rightarrow$ 0\% at $\ell_\infty \le 0.001$ & Showed extreme vulnerability to minimal perturbations. \\
    \addlinespace
    Jati et al. \cite{jati2021adversarial} & 2021 & -- & Deep Speaker & FGSM, PGD, C\&W & Accuracy drops to ~0\% under C\&W & Focuses on digital white-box; highlights PGD-AT as best defense. \\
    \addlinespace
    Sankala et al. \cite{sankala2025adversarial} & 2025 & -- & Text-Dependent SV & Universal Perturbation & 50\%+ (Cross-phrase) & Bypasses passphrase check without victim data; limited to TD-SV. \\
    \midrule
    \multicolumn{7}{l}{\textbf{Black-Box Attacks}} \\
    \midrule
    Zhang et al. \cite{zhang2020voiceprint} & 2020 & Score-based, Stealth & i-vector, x-vector & Psychoacoustic masking (VMask) & FAR: 3.9\% $\rightarrow$ 88.2\% & Perturbations are psychoacoustically masked to be imperceptible. \\
    \addlinespace
    Chen et al. \cite{chen2021real} & 2021 & Score-based, Targeted (FakeBob) & GMM-UBM, i-vector & Natural Evolution Strategy + BIM & ASR: 90.5\% (i-vector) & Effective but requires thousands of queries per sample. \\
    \addlinespace
    Chen et al. \cite{chen2023qfa2sr} & 2023 & Query-free, Transfer-based (QFA2SR) & Commercial APIs (Azure) & Surrogate ensemble + Knowledge Distillation & ASR up to 89.3\% & Eliminates query cost but requires costly surrogate model training. \\
    \midrule
    \multicolumn{7}{l}{\textbf{Physical Attacks}} \\
    \midrule
    Chen et al. \cite{chen2022as2t} & 2022 & Transfer-based (AS2T) & SV, CSI, OSI systems & Multi-task learning + RIR simulation & ASR up to 98.5\% & Demonstrates strong generalization across different speaker recognition tasks. \\
    \addlinespace
    Chen et al. \cite{chen2022push} & 2022 & Real-time (PhyTalker) & Real-time SRS & Time-domain phoneme injection & Targeted ASR: 85\% (OTA) & Real-time physical attack without requiring model access. \\
    
    \addlinespace
    Ahmed et al. \cite{ahmed2023tubes} & 2023 & -- & Speaker Identification & Analog Acoustic (PVC Pipes) & 60\% (Targeted) & Bypasses digital detectors; limited by physical device requirements. \\
        
    \addlinespace
    Xiao et al. \cite{xiao2025timbreadv} & 2025 & -- & Real-time SV & Universal Timbre Perturbation & 95.1\% (Digital), ~50\% (OTA) & Asynchronous and real-time; highly transferable but timbre-focused. \\
    \addlinespace
    \bottomrule
\end{tabular}
}
\end{table*}

\section{Deepfake Attacks}
\label{sec:deepfake_attacks}

Deepfake speech attacks exploit advanced voice synthesis techniques, such as Voice Conversion (VC) and Text-to-Speech (TTS), to impersonate users in Speaker Recognition Systems (SRS). These attacks can bypass both automated authentication systems and human listeners by generating speech that closely mimics a target speaker’s voice, often using minimal training data. Unlike traditional replay spoofing, deepfake attacks can fabricate entirely new utterances, significantly expanding the threat landscape. As generative models become more powerful and accessible, these attacks pose a growing challenge to voice-based security. A summary of key deepfake attacks is presented in Table~\ref{tab:deepfake_attacks}.

\textit{Scope Note:} This survey focuses specifically on studies that evaluate speech synthesis technologies as attack vectors against speaker recognition systems. While the field of generative audio synthesis is vast, encompassing foundational models like Tacotron~2 \cite{shen2018natural} and advanced architectures like VALL-E \cite{wang2023neural}, we exclude works that do not provide an empirical assessment of their models in an adversarial context. This scope ensures our analysis remains centered on demonstrated vulnerabilities in voice biometrics. For a broader overview of modern speech synthesis technologies, see \cite{van2016wavenet,wang2017tacotron,shen2018natural,arik2017deep, wang2023neural, jakubec2024deep, ping2017deep, ren2019fastspeech, ren2020fastspeech, kim2020glow, valle2020flowtron, popov2021grad, kaneko2018cyclegan, kameoka2018stargan, qian2019autovc, chou2019one, qin2023openvoice, peng2024voicecraft, casanova2024xtts, li2025sf, meng2025ds, ji2024controlspeech, li2024dmdspeech, wang2024maskgct, li2023freevc, huang2024diffvc+, fish-speech-v1.4}.

\subsection{Problem Formulation}
In the context of Voice Conversion (VC), the problem is defined as learning a mapping function $G_{A\to B}$ that transforms a source utterance $x_A$ from speaker $A$ into a target utterance $\hat{x}_{A\to B}$ that preserves the linguistic content of $A$ but adopts the timbre of speaker $B$.

We define the generator $G$ and discriminator $D$ operating in a minimax game. The objective function is a weighted sum of adversarial and auxiliary losses:

\begin{equation}
\mathcal{L}_{total} = \mathcal{L}_{adv} + \lambda_{cyc}\mathcal{L}_{cyc} + \lambda_{id}\mathcal{L}_{id}
\end{equation}

Each component is mathematically defined as:
\begin{enumerate}
    \item \textbf{Adversarial Loss ($\mathcal{L}_{adv}$):} Ensures the synthesized audio distribution matches the target speaker's distribution $P_B$.
    \begin{equation}
    \mathcal{L}_{adv} = \mathbb{E}_{x \sim P_B}[\log D(x)] + \mathbb{E}_{x \sim P_A}[\log(1 - D(G(x)))]
    \end{equation}

    \item \textbf{Cycle Consistency Loss ($\mathcal{L}_{cyc}$):} Ensures the transformation is invertible (preserving linguistic content).
    \begin{equation}
    \mathcal{L}_{cyc} = \lVert G_{B\to A}(G_{A\to B}(x_A)) - x_A \rVert_1
    \end{equation}

    \item \textbf{Identity Loss ($\mathcal{L}_{id}$):} Ensures the generator does not alter audio that already belongs to the target speaker.
    \begin{equation}
    \mathcal{L}_{id} = \lVert G_{A\to B}(x_B) - x_B \rVert_1
    \end{equation}
\end{enumerate}

\subsection{Foundational Work}

One of the earliest empirical demonstrations of AI-driven voice spoofing was presented by Mukhopadhyay et al.~\cite{mukhopadhyay2015voices}. Using a commercially available voice morphing tool, they synthesized a target speaker's voice from just a few minutes of enrollment data. The resulting audio was evaluated against both automated speaker verification systems and human listeners. The results were striking: state-of-the-art SRS accepted 80–90\% of the spoofed attempts, while human listeners correctly identified the fakes only about half the time.

This study served as an early warning of the potency of voice conversion attacks and underscored the limitations of both machine-based and human auditory verification. It also highlighted the urgent need for robust anti-spoofing and liveness detection mechanisms as high-fidelity synthesis technologies became more widespread.

\subsection{Partial Fake (PF) Attacks}

Recent research has explored the threat of \textit{partial fake (PF)} speech, where only specific segments of an utterance are synthetically generated while others remain authentic. Alali et al.~\cite{alali2025partial} investigated PF attacks using modern zero-shot TTS and VC models, including VITS and YourTTS. Their findings revealed that both humans and SRS models are highly susceptible to these nuanced manipulations. Because they preserve characteristics of genuine speech, partial fakes are often more challenging to detect than fully synthetic samples, particularly in zero-shot scenarios that require no speaker-specific fine-tuning.

Although the study demonstrates high attack success rates, it also highlights several research gaps. The attacks were not evaluated under noisy or other challenging real-world conditions, such as those involving channel variability or background speech. Consequently, the robustness of these attacks and potential countermeasures in practical deployment scenarios remains an open area for investigation.

\subsection{High-Fidelity and Zero-Shot Attacks}

High-fidelity and zero-shot attacks represent a qualitative shift in the deepfake threat landscape. Unlike earlier voice conversion approaches that required substantial speaker-specific training data, the systems examined here can synthesize a convincing voice profile from only seconds of reference audio, or, in the extreme case, from a single facial image. This dramatic reduction in the data barrier lowers 
the cost of a targeted impersonation attack to the point where it is accessible to non-expert adversaries. The studies below document this transition and reveal that commercially deployed authentication systems, including high-security banking APIs, are already within the reach of these techniques.

Jiang et al. \cite{jiang2024can} demonstrated the ``Foice" attack, a cross-modal deepfake method that synthesizes a target's voice using only a single face image, exploiting the biological correlation between facial structure and vocal characteristics. This approach proved highly effective against commercial systems like WeChat and Microsoft Azure, achieving high bypass rates by leveraging publicly available face data, thus significantly lowering the barrier for large-scale attacks. However, because the method relies on statistical face-voice correlations, it may struggle to capture unique behavioral vocal nuances that are not morphologically determined, limiting its fidelity for specific targets.

Hong et al. \cite{hong2026vulnerabilities} provided an extensive empirical evaluation of modern Voice Conversion (VC) and Text-to-Speech (TTS) models, such as Bert-VITS2 and GPT-SoVITS, against state-of-the-art speaker verification systems. Their findings revealed critical vulnerabilities, with newer models like Bert-VITS2 achieving bypass rates as high as 82.7\% even against systems tuned for high security, significantly outperforming older baselines. A key limitation noted is that while these models excel in digital bypass, the study primarily focuses on the algorithmic capability of the generators and less on their robustness to physical channel distortions or noisy environments.

\begin{table*}[htbp]
    \centering
    \caption{Summary of Key Deepfake Attacks on Voice Authentication Systems categorized by research focus.}
    \label{tab:deepfake_attacks}
    \renewcommand{\arraystretch}{1.3}
    \resizebox{\textwidth}{!}{
    \begin{tabular}{p{2cm}p{0.6cm}p{2.5cm}p{2.2cm}p{3cm}p{2.5cm}p{4.2cm}}
        \toprule
        \textbf{Reference} & \textbf{Year} & \textbf{Attack Type} & \textbf{Target} & \textbf{Method} & \textbf{Key Results} & \textbf{Limitations / Notes} \\
        \midrule
        \multicolumn{7}{l}{\textbf{Foundational Empirical Work}} \\
        \midrule
        Mukhopadhyay et al. \cite{mukhopadhyay2015voices} & 2015 & Full Voice Conversion & ASV \& Human Listeners & Voice morphing using a few minutes of source audio & ASV bypass rate: 80--90\% \newline Human accuracy $\approx$ 50\% & Used early-generation tools; lacked environmental or cross-lingual testing. \\
        \midrule
        \multicolumn{7}{l}{\textbf{Partial Manipulations}} \\
        \midrule
        Alali et al. \cite{alali2025partial} & 2025 & Partial Fake Speech & SRS \& Human Listeners & Zero-shot TTS (YourTTS) and VC (VITS) for partial synthesis & High success rates; reduced detection accuracy for both systems and humans. & Not evaluated in noisy or multilingual settings; defenses not thoroughly assessed. \\
        \midrule
        \multicolumn{7}{l}{\textbf{High-Fidelity \& Zero-Shot Attacks}} \\
        \midrule
        Jiang et al. \cite{jiang2024can} & 2024 & Cross-modal (Foice) & Commercial ASV (WeChat/Azure) & Voice synthesis from single face image & High bypass rates via visual-vocal correlation & Struggles with non-morphological vocal nuances; data privacy concerns. \\
        \addlinespace
        Hong et al. \cite{hong2026vulnerabilities} & 2026 & Modern VC/TTS Attack & High-security ASV systems & Evaluation of Bert-VITS2 and GPT-SoVITS & 82.7\% bypass rate for Bert-VITS2 & Focuses on digital bypass; limited testing on physical channel distortions. \\
        \bottomrule
    \end{tabular}}
\end{table*}

\section{Adversarial Spoofing Attacks}
\label{sec:adv_attacks_cms}

Anti-spoofing countermeasures (CMs), also known as Presentation Attack Detection (PAD), are a critical defense layer in Speaker Verification (SV) pipelines, designed to distinguish bona fide speech from spoofed audio. However, recent studies have demonstrated that these systems are themselves highly vulnerable to adversarial attacks. These attacks introduce crafted, often imperceptible, perturbations to degrade the reliability of CMs, thereby increasing the false acceptance of synthetic or replayed speech. Table~\ref{tab:adversarial_spoofing_attacks} presents a summary of these attacks.

\subsection{Problem Formulation}
Adversarial spoofing is modeled as a Multi-Task Optimization problem. The attacker aims to bypass two distinct systems simultaneously: the Automatic Speaker Verification system ($f_{asv}$) and the Countermeasure system ($f_{cm}$).

Let $x_{spoof}$ be a synthesized audio sample (e.g., from TTS). The attacker seeks a perturbation $\delta$ that minimizes a joint loss function $\mathcal{L}_{joint}$:

\begin{equation}
\min_{\delta} \; \mathcal{L}_{joint} = \mathcal{L}_{asv}(f_{asv}(x_{spoof} + \delta), y_{target}) + \lambda \cdot \mathcal{L}_{cm}(f_{cm}(x_{spoof} + \delta), y_{bonafide})
\end{equation}

where:
\begin{itemize}
    \item $\mathcal{L}_{asv}$ is the impersonation loss (e.g., minimizing distance to target embedding).
    \item $\mathcal{L}_{cm}$ is the evasion loss (e.g., maximizing the probability of the ``bona fide" class).
    \item $\lambda$ is a hyperparameter balancing impersonation capability and spoofing stealth.
\end{itemize}

The attack is considered successful if and only if the joint score satisfies the decision thresholds $\tau_{asv}$ and $\tau_{cm}$ of both systems:

\begin{equation}
\text{Success} \iff S_{asv}(x + \delta) > \tau_{asv} \quad \land \quad S_{cm}(x + \delta) > \tau_{cm}
\end{equation}

\subsection{White-Box Attacks}
White-box adversarial spoofing attacks operate under the assumption that the adversary has access to the gradients of both the speaker verification model and the anti-spoofing countermeasure. This allows joint optimization of a single perturbation that simultaneously satisfies the impersonation objective and evades spoof detection, a dual-target problem that dramatically narrows the space of viable defenses.

Early work by Liu et al.~\cite{liu2019adversarial} applied white-box adversarial techniques, such as the FGSM and PGD, against PAD models. Even minimal, imperceptible perturbations (i.e., small $\epsilon$) increased detector error rates to as high as 85–95\%. These results revealed that, with access to model gradients, an attacker can nearly eliminate a PAD system's ability to detect spoofed speech, effectively reducing its performance to that of random guessing.

Building on this, Gomez-Alanis et al.~\cite{gomez2022adversarial} proposed a trainable adversarial generator named ABTN (Adversarial Biometrics Transformation Network), designed to optimize perturbations using both PAD and Automated Speaker Verification (ASV) losses. ABTN outperformed traditional FGSM by generating perturbations that generalized better across spoof types while preserving the speaker’s identity. This allowed the attack to fool the PAD system without compromising the ASV system's ability to verify the speaker. The method improved the attack Equal Error Rate (EER) by an absolute 27–51\% on the ASVspoof 2019 logical and physical access protocols \cite{Todisco2019ASVspoof}.

A different approach, GANBA~\cite{gomez2022ganba}, introduced a conditional Generative Adversarial Network (GAN) architecture capable of simultaneously fooling both the SV and PAD modules. Trained on bona fide and spoofed samples, the model generated high-quality synthetic utterances that significantly degraded system performance. Evaluations on ASVspoof 2019 showed that SV accuracy dropped below 20\% while PAD effectiveness was severely compromised, illustrating the threat posed by jointly optimized adversarial audio.

\subsection{Black-Box Attacks}
To overcome the limitations of white-box assumptions, several studies have explored black-box strategies that do not require gradient access.

Xuan Hai et al.~\cite{hai2023sifdetectcracker} presented \textit{SiFDetectCracker}, a query-based black-box attack that perturbs non-speech segments (e.g., silence, background noise) to exploit the reliance of some PAD systems on environmental cues. The attack achieved a success rate of over 80\% without altering speaker-identifying vocal components, ensuring the audio was still accepted by the ASV system. This highlights that adversarial success does not require manipulating the entire signal—targeting peripheral acoustic features can be sufficient.

A more physically robust approach was introduced by Kassis et al.~\cite{kassis2023breaking}, who proposed a lightweight signal processing attack that successfully deceived both ASV and CM systems during over-the-air playback. By cloning prosodic and spectral features from high-quality speaker recordings, the attack bypassed multiple detection systems. However, it was less effective against advanced models, such as Wav2Vec2-based PADs \cite{baevski2020wav2vec}. The attack also assumes that the attacker can inject audio via a rooted phone, a prerequisite often blocked by the security measures of high-stakes applications \cite{kassis2023breaking}.

Another category of black-box attacks involves creating universal adversarial filters. Panariello et al.~\cite{panariello2023malafide} introduced \textit{Malafide}, a fixed Finite Impulse Response (FIR) filter learned via adversarial training. This universal filter can be convolved with any spoofed utterance to evade detection, removing the need to craft input-specific perturbations. This method proved effective against high-performing deep PADs and is practical for real-time applications.

This approach was extended in \textit{Malacopula}~\cite{todisco2024malacopula}, which created a nonlinear filter targeting the ASV component itself by enhancing spoofed audio to better match the victim’s voiceprint. These methods demonstrate a clear trend toward physically realizable adversarial strategies that can survive digital-to-analog conversion and transmission through real-world channels.

Kamel et al. \cite{kamel2025spectral} introduced the Spectral Masking and Interpolation Attack (SMIA), a black-box adversarial framework designed to bypass combined voice authentication and anti-spoofing systems. By strategically manipulating the frequency regions of AI-generated audio that are inaudible to the human ear, SMIA generates adversarial samples that preserve perceptual quality while successfully deceiving security models. The attack achieved an 82\% success rate in bypassing joint defenses, demonstrating that attackers can exploit spectral redundancy to hide adversarial traces. However, because the attack relies on manipulating specific frequency bands often considered ``inaudible," its effectiveness may be reduced by lossy compression codecs or band-limited transmission channels (e.g., telephony) that filter out these non-essential frequencies.

Hai et al. \cite{hai2025sifmimicevader} introduced ``SiFMimicEvader," a black-box adversarial attack framework that evades deepfake detectors by training a surrogate ``mimic" detector to approximate the target's behavior. This method leverages Speaker-irrelative Features (SiF) to craft perturbations that camouflage synthetic audio as real, achieving an attack success rate of over 90\% against top detectors like RawNet2 and AASIST without needing direct access to the target model's gradients. However, the attack's success is partly dependent on the transferability between the mimic and the target, and performance can vary if the target architecture is significantly different from the surrogate.

\begin{table*}[htbp]
\centering
\caption{Summary of Adversarial Spoofing Attacks on Voice Authentication and Anti-Spoofing Systems.}
\label{tab:adversarial_spoofing_attacks}
\resizebox{\textwidth}{!}{
\begin{tabular}{p{2cm}p{0.7cm}p{1.7cm}p{3.2cm}p{1.4cm}p{3.5cm}p{3.5cm}}
    \toprule
    \textbf{Reference} & \textbf{Year} & \textbf{Access} & \textbf{Method} & \textbf{Target} & \textbf{Effect on Target} & \textbf{Remarks / Limitations} \\
    \midrule
    \multicolumn{7}{l}{\textbf{White-Box Attacks}} \\
    \midrule
    Liu et al. \cite{liu2019adversarial} & 2019 & White-box & Gradient-based (FGSM, PGD) & PAD & Increases error rate to $>$85\% & Requires gradient access. \\
    \addlinespace
    Gomez-Alanis et al. \cite{gomez2022adversarial} & 2022 & White-box & ABTN generator & PAD, ASV & +27--51\% EER (absolute) & Generalizes across spoof types. \\
    \addlinespace
    Gomez-Alanis et al. \cite{gomez2022ganba} & 2022 & White-box & GAN-based generation & PAD, ASV & ASV accuracy $<$20\% & Computationally intensive. \\
    \midrule
    \multicolumn{7}{l}{\textbf{Black-Box Attacks}} \\
    \midrule
    Hai et al. \cite{hai2023sifdetectcracker} & 2023 & Black-box & Non-speech perturbation & PAD & $>$80\% success rate & Query-heavy; tailored to silence. \\
    \addlinespace
    Kassis et al. \cite{kassis2023breaking} & 2023 & Black-box & Signal processing / Playback & ASV, CM & Physical bypass & Less effective on Wav2Vec2. \\
    Panariello et al. \cite{panariello2023malafide} & 2023 & Black-box & Universal FIR filter & PAD & Real-time universal bypass & Specific to target model. \\
    \addlinespace
    Todisco et al. \cite{todisco2024malacopula} & 2024 & Black-box & Universal nonlinear filter & ASV & Enhanced voiceprint match & Physically viable; targets ASV. \\
    \addlinespace
    Kamel et al. \cite{kamel2025spectral} & 2025 & Black-box & Spectral Masking (SMIA) & PAD + ASV & 82\% bypass rate & Vulnerable to lossy compression. \\
    \addlinespace
    Hai et al. \cite{hai2025sifmimicevader} & 2025 & Black-box & Surrogate Mimic Detector & PAD & $>$90\% ASR (RawNet2/AASIST) & Success depends on transferability. \\
    \bottomrule
\end{tabular}
}
\end{table*}

\section{Evaluation}
\label{sec:evaluation}

Robust evaluation is critical for measuring the effectiveness of both attack strategies and defense mechanisms in a reproducible manner. This section details the standard datasets used to benchmark performance, defines the core metrics for assessment, and summarizes the key competitions that have driven recent advancements in the field.

\subsection{Popular Datasets}
Evaluating VAS and studying their vulnerabilities requires access to corpora that span a wide spectrum of acoustic complexity. Over the past decade, the field has transitioned from small-scale, controlled environments to massive ``in-the-wild'' datasets necessary for deep learning generalization. A detailed comparison of dataset metrics, including size, diversity, and primary application is provided in Table~\ref{tab:dataset_metrics_diff_final}.

Foundational, or ``legacy,'' corpora such as \textbf{YOHO}~\cite{campbell1995testing} and \textbf{TIMIT}~\cite{Garofolo1993TIMIT} remain vital for phonetic alignment and text-dependent verification but offer low acoustic diversity due to their controlled lab environments. \textbf{NTIMIT}~\cite{jankowski1990ntimit}, and portions of \textbf{LibriSpeech}~\cite{Panayotov2015LibriSpeech} introduce necessary channel 
variability through telephone-band limitations and varying recording qualities, serving as baselines for channel robustness.

In contrast, modern SV research relies on high-entropy, unconstrained datasets. \textbf{VoxCeleb1/2}~\cite{Nagrani2017VoxCeleb, chung2018voxceleb2} have become the industry standard, providing thousands of hours of speech that challenge models with overlapping voices, background noise, and codec artifacts inherent to YouTube audio. Similarly, Mozilla's \textbf{Common Voice}~\cite{CommonVoice} offers extreme 
hardware diversity ($>$28,000 hours recorded on diverse consumer devices), while \textbf{AISHELL-2}~\cite{Yu2020AISHELL2} specifically introduces mobile-channel distortions (e.g., iOS recording characteristics) for Mandarin speech.

In the adversarial domain, the definition of ``difficulty'' has shifted from acoustic artifacts to logical consistency. The \textbf{ASVspoof} 
challenges~\cite{Wu2015ASVspoof, Kinnunen2017ASVspoof, Todisco2019ASVspoof, Yamagishi2021ASVspoof} track this evolution, progressing from detecting vocoder ``buzziness'' (2015) to identifying Neural TTS and Deepfakes masked by compression and VoIP transmission (2021). New benchmarks like \textbf{Fake or Real (FoR)}~\cite{Zhang2021FoR} provide rigorously balanced subsets to prevent classifiers from learning normalization cues, while \textbf{BackdoorVoice}~\cite{Qi2022BackdoorVoice} addresses the security threat of data poisoning, where models are compromised by imperceptible triggers rather than overt spoofing. Finally, the \textbf{SVCC} datasets~\cite{LorenzoTrueba2020SVCC} 
introduce the high-difficulty task of singing voice conversion, requiring models to disentangle melodic dynamics from speaker timbre.

\begin{table*}[h]
\centering
\caption{Comparison of Voice Authentication and Spoofing Datasets.}
\label{tab:dataset_metrics_diff_final}
\renewcommand{\arraystretch}{1.3}
\resizebox{\textwidth}{!}{%
\begin{tabular}{@{}llcllllc@{}}
    \toprule
    \textbf{Dataset} & \textbf{Main Application} & \textbf{\# Spkrs} & 
    \textbf{Size} & \textbf{Diversity} & \textbf{Difficulty} & 
    \textbf{Content Type} & \textbf{Spoofing Attacks} \\ \midrule
    \textbf{YOHO} & Speaker Verification & 138 & $\sim$20--25 hrs & 
    Low & Low & Fixed phrases & — \\
    \addlinespace
    \textbf{VoxCeleb1/2} & Speaker Verification & 1.2k / 6.1k & 
    $\sim$350 / 2,400 hrs & Med / High & Med / High & Celebrity interviews & — \\
    \addlinespace
    \textbf{AISHELL-1/2} & Speaker Verification & 400 / 1,991 & 
    178 / 1,000 hrs & Low / Med & Low / Med & Mandarin read speech & — \\
    \addlinespace
    \midrule
    \textbf{TIMIT} & Speaker Verification & 630 & 5.4 hrs & 
    Low & Low & Scripted studio speech & — \\
    \addlinespace
    \textbf{NTIMIT} & Channel Robustness & 630 & $\sim$5 hrs & 
    Med & Med & Telephone-quality & — \\
    \addlinespace
    \textbf{LibriSpeech} & Speaker Verification & 2,484 & $\sim$1,000 hrs & 
    Med & Med & Audiobooks & — \\
    \addlinespace
    \textbf{Common Voice} & Speaker Verification & $>$90,000 & 
    $>$28,000 hrs & Very High & High & Crowdsourced & — \\
    \addlinespace
    \midrule
    \textbf{VCTK} & Voice Conversion & 109 & $\sim$44 hrs & 
    Med & Low & Read speech, accented & — \\
    \addlinespace
    \textbf{SVCC} & Singing Voice Conversion & Var. & Var. ($\sim$10 min) & 
    High & Very High & Singing voice & VC \\
    \addlinespace
    \midrule
    \textbf{ASVspoof 2015} & Anti-Spoofing & 106 & $\sim$260k trials & 
    Low & Low & Logical (TTS/VC) & Synth., Replay \\
    \addlinespace
    \textbf{ASVspoof 2017} & Anti-Spoofing (Replay) & 10 & $\sim$40 hrs & 
    Med & Med & Replay detection & Replay \\
    \addlinespace
    \textbf{ASVspoof 2019} & Anti-Spoofing & 121 & $\sim$250 hrs & 
    High & High & Logical \& Physical & VC, TTS, Replay \\
    \addlinespace
    \textbf{ASVspoof 2021} & Anti-Spoofing & 208 & Millions of trials & 
    Very High & Very High & DF, Codec, Noisy & Neural VC, TTS, Replay \\
    \addlinespace
    \textbf{FoR} & Deepfake Detection & 110 & 195,000 utts & 
    High & High & Real vs. TTS & TTS \\
    \midrule
    \textbf{BackdoorVoice} & Backdoor / Poisoning & 150 & Variable & 
    High & High & Triggered utterances & Backdoor \\ 
    \bottomrule
\end{tabular}%
}
\end{table*}

\subsection{Evaluation Metrics}
The standard metrics for evaluating SV systems include:
\begin{itemize}
  \item \textbf{Equal Error Rate (EER)}: The rate at which the false acceptance rate (FAR) equals the false rejection rate (FRR). Lower EER indicates better system accuracy.
  
  \item \textbf{Detection Error Tradeoff (DET) Curve}: A plot showing the tradeoff between FAR and FRR over different thresholds, used to visualize system performance.

  \item \textbf{Attack Success Rate (ASR)}: The proportion of adversarial attempts that successfully fool the system. A higher ASR indicates the system is more vulnerable.

  \item \textbf{Signal-to-Noise Ratio (SNR)}: Measures the strength of the original signal relative to adversarial noise. Higher SNR indicates less perceptible attacks.\\
  $\text{SNR (dB)} = 10 \log_{10} \left( \frac{P_{\text{signal}}}{P_{\text{noise}}} \right)$

  \item \textbf{t-DCF (tandem Detection Cost Function)}: A primary metric for evaluating automatic speaker verification systems in combination with countermeasure systems. Lower t-DCF values indicate better joint system performance.\\
  $\text{t-DCF} = C_{\text{miss}}^{\text{ASV}} \cdot \pi_{\text{tar}} \cdot P_{\text{miss}}^{\text{ASV}} + C_{\text{fa}}^{\text{ASV}} \cdot \pi_{\text{non}} \cdot P_{\text{fa}}^{\text{ASV}} + C_{\text{fa}}^{\text{CM}} \cdot \pi_{\text{spoof}} \cdot P_{\text{fa}}^{\text{CM}} + C_{\text{miss}}^{\text{CM}} \cdot \pi_{\text{tar}} \cdot P_{\text{miss}}^{\text{CM}}$
\end{itemize}

\subsection{Competitions}
Table~\ref{tab:anti-spoofing} provides a comparative overview of the major anti-spoofing competitions, summarizing their key characteristics, attack types, and contributions to the field.

Several benchmark competitions have significantly improved the field of \textbf{speaker verification}. \textbf{The NIST Speaker Recognition Evaluations (SRE)}~\cite{nist_sre} have shaped speaker recognition research since 1996, benchmarking systems under demanding real-world conditions such as language variability, channel effects, and environmental noise.

More recently, the \textbf{VOiCES from a Distance Challenge} \cite{voices_challenge} has targeted speaker recognition in far-field and acoustically challenging environments, using data captured in realistic room settings. Such challenges push the boundaries of SRS in practical, low-SNR conditions.

With the rise of spoofing threats, dedicated \textbf{anti-spoofing competitions} have emerged. The \textbf{ASVspoof Challenge series} \cite{Wu2015ASVspoof, Kinnunen2017ASVspoof, Todisco2019ASVspoof, Yamagishi2021ASVspoof} is the most prominent, providing standardized protocols and datasets to evaluate spoof detection systems. Each edition of ASVspoof introduces new attack types, evolving from replay and voice conversion to advanced synthetic speech generated by neural vocoders and TTS systems.

Other initiatives include the \textbf{Microsoft ADD Challenge} \cite{add2022}, which broadens the scope to deepfake audio detection, and the \textbf{Deepfake Detection Challenge (DFDC)} \cite{dfdc2020}, which, although primarily visual, includes audio-based manipulations in its multi-modal dataset.

\begin{table*}[htbp]
\centering
\caption{Summary of Major Anti-Spoofing Competitions.}
\label{tab:anti-spoofing}
    \resizebox{\textwidth}{!}{ 
    \begin{tabular}{lcccp{5.5cm}}
    \toprule
    \textbf{Competition} & \textbf{Year} & \textbf{Attacks} & \textbf{Focus} & \textbf{Key Contribution} \\
    \midrule
    ASVspoof 2015 \cite{Wu2015ASVspoof} & 2015 & VC, SS & Standalone ASV spoofing & Introduced standard protocols for synthetic spoofing detection \\
    \addlinespace
    ASVspoof 2017 \cite{Kinnunen2017ASVspoof} & 2017 & Replay & Physical replay attacks & Real-world replay attacks recorded in diverse acoustic conditions \\
    \addlinespace
    ASVspoof 2019 \cite{Todisco2019ASVspoof} & 2019 & VC, SS, Replay & Logical and physical access & Used neural vocoders for synthesis; first division into LA and PA tasks \\
    \addlinespace
    DFDC Audio Track \cite{dfdc2020} & 2020 & Deepfake (multi-modal) & Multimodal fake detection & Included manipulated audio in conjunction with manipulated video \\
    \addlinespace
    ASVspoof 2021 \cite{Yamagishi2021ASVspoof} & 2021 & LA, DF, VC & Generalized anti-spoofing & Focused on generalization across spoofing types and unseen attacks \\
    \addlinespace
    Microsoft ADD \cite{add2022} & 2022 & TTS, VC, Deepfake & Audio deepfake detection & Realistic multi-accent dataset targeting fake audio from TTS/VC \\
    \bottomrule
    \end{tabular}
    }
\end{table*}

\section{Real-World Deployment and Challenges}
\label{sec:realworld}

\textbf{A note on sources.} Production deployment data for voice authentication is largely undocumented in the peer-reviewed literature: commercial banks and contact-center operators rarely publish internal audit results, and academic case studies of live systems are scarce. We therefore rely on practitioner sources—such as industry press (Forbes, Vice, Banking Dive), market-research summaries (Mordor Intelligence, MarketsandMarkets), and vendor and regulatory communications—when peer-reviewed evidence is unavailable. We treat these as \textit{grey literature}: useful for establishing what is being deployed and what incidents have been reported, but not as primary evidence of attack effectiveness. When peer-reviewed work or official regulatory documents exist (e.g., on GDPR/BIPA frameworks, EBA payment services guidance, and consumer acceptance surveys), we cite them alongside the practitioner sources.

Voice authentication systems have moved rapidly from research laboratories into production environments across critical industries. However, this widespread adoption has surfaced a set of challenges that extend well beyond the attack landscape analyzed in Sections~\ref{sec:data_poisoning}--\ref{sec:adv_attacks_cms}. We organize these challenges into four dimensions. \textbf{First, security challenges}, which are dominated by the attacks discussed throughout this survey, constitute the \textit{major technical challenge} facing deployed systems and are revisited below from a deployment perspective rather than a methodological one. \textbf{Second, operational challenges} arising from real-world acoustic conditions (background noise, network variability, illness-induced voice changes) that degrade accuracy and force fallback to weaker authentication. \textbf{Third, regulatory and privacy challenges} stemming from the immutability of biometric voiceprints under GDPR, BIPA, and emerging payment services directives. \textbf{Fourth, user-acceptance challenges}, the trust deficit and usability friction that limit adoption. The remainder of this section examines each dimension, and Section~\ref{sec:comparison} then summarizes the cross-cutting research-methodology challenges that affect both attack and defense studies.

\subsection{Industry Adoption and Current Landscape}

The financial services sector has led the charge in deploying voice biometrics, driven primarily by fraud prevention concerns and the need to secure telephone banking channels. Major U.S. institutions, including JPMorgan Chase~\cite{chase2024voiceid, finextra2018jpmorgan}, Wells Fargo~\cite{wellsfargo2024voice, bankingdive2019biometrics}, and TD Bank~\cite{biometricupdate2023senate}, have integrated voice authentication into their customer service operations. Internationally, the Bank of Ireland's €34 million investment in 2024 exemplifies the scale of institutional commitment, with voice identification systems now processing approximately 11,000 customer calls daily while targeting both fraud reduction and operational efficiency gains~\cite{bankofireland2024, biometricupdate2024bankofireland}.

This technology has also gained traction in telecommunications, government services, healthcare patient verification, and retail contact centers---all domains characterized by high transaction volumes and significant fraud exposure~\cite{mordor2025voice, marketsandmarkets2024voice}. The appeal is straightforward: voice authentication eliminates the friction of remembering passwords or PINs while offering what appears to be a robust biometric credential. Organizations report measurable reductions in call handling times and decreased dependence on human verification processes, as voiceprints derived from hundreds of vocal characteristics replace traditional knowledge-based authentication~\cite{fortunebi2024voice}.

\subsection{Security Challenges and the Deployment Disconnect}

Despite this momentum, a troubling gap has emerged between commercial deployments and their ability to withstand contemporary threats, particularly those leveraging advances in artificial intelligence.

\textbf{Vulnerability to Synthetic Speech.} Independent security audits have revealed that commercially deployed banking voice authentication systems can be successfully compromised using synthetic speech from accessible voice-cloning tools~\cite{vice2024voiceai, linkedin2020banks}. These are not hypothetical attacks---penetration testing has demonstrated that high-fidelity deepfakes can bypass operational systems protecting real financial accounts~\cite{vice2024voiceai}. The deepfake threats discussed in Section~\ref{sec:deepfake_attacks} have transitioned from academic concerns to active security risks, yet many deployed systems rely on countermeasure techniques that struggle to keep pace with rapidly evolving synthesis methods~\cite{brown2024scams, ny12024fraud}.

\textbf{Real-World Operating Conditions.} Laboratory benchmarks rarely capture the complexity of production environments. Deployed voice authentication systems must handle background noise, variable network quality, and natural variations in users' voices caused by illness, stress, or fatigue. These factors often degrade recognition accuracy, leading to higher false-rejection rates~\cite{markntel2024voice}. When legitimate users are locked out, systems typically revert to traditional authentication methods---PINs, security questions, or agent verification---thereby reintroducing the vulnerabilities that voice biometrics were intended to eliminate~\cite{bankingexchange2016biometrics}.

\subsection{Regulatory Pressure and Privacy Concerns}

The biometric nature of voice data has drawn regulatory scrutiny. Unlike passwords, which can be changed if compromised, voiceprints represent permanent physiological characteristics. This immutability places voice data in the highest tier of sensitivity under frameworks such as the EU's General Data Protection Regulation (GDPR)~\cite{gdprlocal2025biometric, ambiq2025voice} and Illinois's Biometric Information Privacy Act (BIPA)~\cite{bloomberglaw2024bipa, imprivata2024bipa, wikipediabipa2026}, triggering requirements for explicit consent, robust encryption, and transparent retention policies.

U.S. Senate committees have questioned whether financial institutions adequately inform consumers about voice data collection and storage practices~\cite{brown2023senate, biometricupdate2023senate, techpolicypresss2024senate}. More fundamentally, European regulators have challenged whether voice authentication alone constitutes ``strong customer authentication'' under payment services directives, citing susceptibility to spoofing~\cite{fraud2023eba, dlapiper2019eba, eba2023wallets}. This regulatory skepticism is driving a shift toward mandatory multi-factor authentication, where voice serves as one component rather than a standalone credential~\cite{bankingexchange2016biometrics}.

\subsection{The User Acceptance Challenge}

Consumer attitudes reveal a paradox. While approximately 81\% of users appreciate the convenience of voice authentication~\cite{iproov2025statistics, mordor2025voice}, 60--70\% harbor concerns about data misuse and security breaches~\cite{iproov2025statistics, getapp2024biometrics}. This trust deficit is amplified by high-profile reports of successful deepfake attacks~\cite{vice2024voiceai, brown2024scams}, which erode confidence in the technology's reliability.

Practical usability issues compound these concerns. Many users feel uncomfortable speaking authentication phrases in public spaces or open office environments. Temporary voice changes from colds, allergies, or vocal strain can trigger authentication failures, creating frustration and diminishing the perceived benefits~\cite{markntel2024voice}. These friction points, combined with lingering security doubts, continue to constrain adoption rates and limit the contexts in which voice authentication can be effectively deployed~\cite{aware2025consumer, globenewswire2024voice}.

\section{Comparison and Critical Analysis}
\label{sec:comparison}

This section presents a comprehensive comparative analysis of the threat vectors examined in the preceding sections, synthesizing their characteristics to identify systemic vulnerabilities within the voice authentication pipeline. Rather than evaluating data poisoning, adversarial perturbations, deepfakes, and adversarial spoofing in isolation, this analysis characterizes them as a continuum of risks that exploit distinct stages of the authentication process. The following discussion contrasts these threats across three critical dimensions: operational feasibility, physical robustness, and the growing asymmetry between offensive capabilities and defensive readiness.

\subsection{Operational Feasibility and The Access-Impact Trade-off}

A critical divergence exists between the \textit{theoretical} severity of an attack and its \textit{practical} feasibility.

\textbf{Knowledge Constraints in Optimization-Based Attacks.} 
    Optimization-based adversarial attacks, such as PGD and FGSM, consistently achieve high ASR, often exceeding 95\% in controlled digital settings. Despite their effectiveness, these methods are subject to significant practical limitations, as they typically require white-box access to model parameters or gradients. This requirement makes them largely infeasible against real-world closed-source systems, including commercial banking IVR platforms and consumer smart speakers. Although black-box extensions attempt to relax this assumption through query-based optimization, they incur substantial computational overhead and often rely on a large number of queries, which increases the likelihood of detection through rate-limiting or monitoring mechanisms.
    
\textbf{Accessibility of Deepfake-Based Attacks.} 
    In contrast, deepfake-based attacks present a more accessible and scalable threat model. VC and TTS techniques operate independently of the internal architecture of the target system, enabling zero-shot attacks that do not require prior knowledge of the downstream voice authentication or verification model. As discussed in Section~\ref{sec:deepfake_attacks}, recent advances allow convincing voice replication from only a few seconds of reference audio. As a result, attacks that are comparatively easy to deploy are increasingly capable of producing outputs that are difficult to distinguish from genuine speech, thereby posing a significant and immediate risk to real-world deployments.

This constraint in optimization-based attacks is structural: gradient-based methods are fundamentally tied to the optimization landscape of a specific model instance, making transferability and physical robustness antagonistic objectives, improving one tends to weaken the other.

\subsection{Digital-to-Physical Generalization and Robustness}

A key limitation across much of the existing literature is the limited consideration of physical deployment conditions. The majority of prior work evaluates attack performance in purely digital environments, without fully modeling the non-linear distortions introduced by microphones, room acoustics, and over-the-air signal propagation.

\textbf{Sensitivity of Adversarial Perturbations.} 
    Optimization-based adversarial perturbations exhibit a high degree of sensitivity to signal degradation during physical playback and recording. Although RIR-aware methods such as AS2T \cite{chen2022as2t} and PhyTalker \cite{chen2022push} attempt to account for room impulse responses, their effectiveness frequently diminishes under realistic conditions. Even minor temporal misalignment, background noise, or device-specific distortions can disrupt the precise perturbation structure required to mislead the target model. This fragility arises because adversarial perturbations operate at a scale ($\ell_\infty < 0.002$) where even sub-millisecond timing jitter during analog-to-digital conversion can shift the signal beyond the perturbation's effective region.

\textbf{Physical Robustness of Synthesis-Based Attacks.} 
    Deepfake and replay attacks, which operate on higher-level semantic characteristics rather than imperceptible signal perturbations, generally exhibit greater robustness to over-the-air transmission. Nevertheless, these approaches often introduce distinctive spectral artifacts, such as vocoder-related distortions, that may be amplified by physical recording devices. Consequently, such artifacts can be more easily detected by artifact-based countermeasures than in purely digital injection settings. The higher physical robustness of deepfake attacks follows from their operating on semantic rather than signal-level representations: as long as the synthesized voice preserves the target speaker's formant structure, the semantic content survives the acoustic distortions introduced by recording equipment.

\subsection{Defense Asymmetry: Optimization vs. Generation}
Our survey reveals a fundamental asymmetry in how defenses respond to different attack paradigms.

\textbf{Reactive vs. Proactive.} Defenses against adversarial spoofing are largely reactive. Systems demonstrate that once an attacker optimizes against a specific Countermeasure (CM), the defense collapses. The current ``cat-and-mouse" dynamic heavily favors the attacker, who only needs to find one optimization path to minimize the joint loss.
\textbf{The Poisoning Paradox.} Data poisoning defenses face a unique paradox. Most detection methods (e.g., spectral clustering or anomaly detection) assume the existence of a ``clean" validation set to calibrate their baselines. In real-world web-scraping scenarios—where models ingest terabytes of uncurated data—guaranteeing this clean baseline is often impossible, leaving models vulnerable to ``clean-label" attacks that are mathematically indistinguishable from natural outliers.

\subsection{Attack Interaction and Compounding Threats}

A critical dimension absent from much of the existing literature is the possibility of \textit{compound attacks}, in which multiple threat vectors are combined into a single unified offensive strategy. Two interaction patterns emerge from our survey.

\textbf{Poisoning as an enabler for adversarial attacks.} If an adversary can first poison the training data to shift the model's decision boundary toward a specific region, the perturbation budget required by a subsequent adversarial attack at inference time is substantially reduced. A backdoor trigger planted during training effectively pre-aligns the model to accept a specific adversarial pattern, transforming what would otherwise require expensive white-box optimization into a trivial inference-time activation. This compounding has not been empirically evaluated in the voice authentication domain, but is a plausible and underexplored threat vector.
\textbf{Deepfake output as adversarial spoofing input.} The pipeline of deepfake generation followed by adversarial perturbation represents the most practically threatening compound attack: the deepfake bypasses semantic identity checks by mimicking the target speaker, while the adversarial perturbation neutralizes the anti-spoofing countermeasure. As shown by \cite{gomez2022ganba} and \cite{hai2025sifmimicevader}, joint optimization of these two objectives is already achievable. What remains unaddressed is whether the interaction is synergistic, each component being more effective within the pipeline, or merely additive. These compound scenarios represent the most realistic adversarial setting for deployed systems and should be the primary focus of future red-teaming evaluations.

\subsection{Cross-Cutting Research Challenges}
\added{Despite the diversity of the attack vectors discussed, several common limitations continue to restrict the applicability of existing research to realistic threat models. The first is a reliance on static evaluation: many frameworks are built on pre-recorded corpora such as VoxCeleb~\cite{Nagrani2017VoxCeleb} and rarely account for the dynamics of real interactions---speaker interruptions, variations in speech rate, or shifts in emotional state---even though these factors materially affect both attack effectiveness and defense performance. The second is limited cross-model and cross-domain generalization: attacks and defenses alike tend to be tightly coupled to particular architectures and data distributions, so that poisoning designed for ResNet-based systems often fails to transfer to x-vector models, and deepfake detectors trained largely on English degrade in cross-lingual use. The third is the burden of real-time operation: optimization-based attacks that succeed offline are frequently too computationally heavy to mount within the latency budget of live communication, leaving their practical feasibility during, for example, an ongoing telephone call an open engineering question.}

\begin{table*}[h]
\centering
\caption{Comparison of representative attack categories against voice authentication systems. Feasibility reflects practical deployment effort; Physical robustness denotes resilience under over-the-air (OTA) conditions; Stealth refers to perceptual detectability by human listeners.}
\label{tab:comparison_matrix}
\resizebox{\textwidth}{!}{%
\begin{tabular}{lccccc}
    \toprule
    \textbf{Attack Category} & \textbf{Core Mechanism} & \textbf{Attacker Knowledge} & \textbf{Deployment Feasibility} & \textbf{Physical Robustness} & \textbf{Perceptual Stealth} \\ 
    \midrule
    Data Poisoning      & Training data manipulation    & White/Grey-box   & Low   & Not applicable    & N/A \\
    Adversarial Attacks     & Gradient-based perturbation   & White-box     & Low   & Very low  & High \\
    Deepfake Attacks    & Generative modeling (VC/TTS)  & Black-box     & High  & Moderate  & Low \\
    Adversarial Spoofing    & Joint attack and CM optimization  & Grey/White-box   & Moderate  & Low   & Moderate \\
    \bottomrule
\end{tabular}%
}
\end{table*}

\section{Future Work}
\label{sec:future_work}

\added{Despite considerable progress in securing voice-based biometric systems, the vulnerabilities surveyed in this paper continue to outpace the defenses meant to contain them. We organize the directions below around the four threat families that structure this survey---data poisoning, adversarial attacks, deepfake synthesis, and adversarial spoofing---followed by a set of challenges that recur across all of them and a discussion of the paradigms we expect to dominate the next phase of research. Throughout, we favor concrete and measurable objectives over general aspirations, since the field's progress has too often been obscured by evaluations that are difficult to compare.}

\subsection{Data Poisoning Attacks}
\added{As poisoning strategies mature from crude label flipping toward clean-label and backdoor formulations, the central difficulty shifts from removing obviously corrupted samples to detecting manipulations that are, by construction, statistically indistinguishable from natural data. The most pressing need is therefore reliable detection of stealthy, low-rate poisoning: a practical target is to identify clean-label infections at rates as low as $0.5\%$ without inflating the false rejection rate by more than one percentage point on standard benchmarks. This is hard precisely because clean-label samples resemble benign outliers under conventional hygiene checks, so detection must fall back on influence-function analysis, which scales poorly to large models, or on embedding-space anomaly detection, which presupposes a guaranteed-clean reference set---a circular requirement in web-scraped training pipelines.}

\added{Equally important is the move toward realistic threat models. Much of the literature still assumes white-box access, whereas the more consequential setting is federated training, where an adversary controlling fewer than ten percent of non-IID client nodes can degrade both convergence time and final equal error rate. Robust aggregation rules such as Krum or trimmed-mean offer limited protection here, because malicious updates are easily confused with those of underrepresented but legitimate speakers, and discarding them penalizes exactly the long-tail population that voice biometrics already serves poorly. Progress on this front depends on understanding how poisoning effects persist---or fail to persist---across architectures and datasets, a transferability question that remains largely unexamined yet ultimately determines whether a single poisoned corpus can compromise many downstream systems. Attention should also turn to the cost and stealth of the attacks themselves: triggers engineered to remain inaudible, for instance above a $40$~dB signal-to-noise ratio, and hybrid pipelines that combine training-time poisoning with inference-time perturbations, whose compound behavior is not captured by defenses designed for either threat in isolation.}

\subsection{Adversarial Attacks}
\added{Although white-box adversarial attacks achieve high success rates in digital simulation, their translation to operational conditions remains the defining open problem. Perturbations rarely survive changes of architecture, dataset, or acoustic environment, and closing this gap will require universal, transferable perturbations---plausibly obtained through meta-learning or training-agnostic optimization---rather than attacks re-crafted for each target. Their evaluation must likewise become physically honest: robustness should be demonstrated across variable room impulse responses, with reverberation times ($T_{60}$) spanning roughly $200$ to $800$~ms, and an attack that sustains success above $80\%$ under these conditions would mark a genuine advance. The obstacle is one of scale, since an adversarial signal whose effective margin is smaller than the distortion introduced by a single unmodeled reflection can be made robust only through expectation-over-transformation across thousands of impulse responses, which is prohibitively expensive when every attack is re-optimized per target.}

\added{Two further constraints separate laboratory results from deployable ones. Real-time interception of streaming voice calls demands perturbations generated with algorithmic latency below roughly $20$~ms, while attacks against rate-limited commercial APIs must be query-efficient, ideally exceeding $90\%$ targeted success with fewer than a thousand queries. The latter is bounded by information theory: hard-label gradient estimation costs grow with input dimensionality, and waveforms carry on the order of sixteen thousand samples per second, so sub-thousand-query attacks depend on strong priors---surrogate gradients or learned attack policies---whose transfer to closed systems is itself unproven. Underlying all of these is the persistent fragility of adversarial performance outside English data, a cross-lingual gap that has received little systematic attention and that will require language-agnostic formulations if the threat is to be characterized globally.}

\subsection{Deepfake Attacks}
\added{The collapse in the cost of high-fidelity voice cloning has shifted the burden decisively onto detection, which must now operate under conditions that current benchmarks largely ignore. Detectors are typically trained and evaluated on clean data, yet they will be deployed amid telephony bandwidth limits, reverberation, crosstalk, and overlapping speech, and their robustness under these degradations is the first thing that must be established. A related and equally neglected axis is linguistic generalization: because most systems are trained on English, a credible objective is zero-shot detection on unseen languages that holds the equal-error-rate gap below five percent relative to seen languages. Where detection does succeed, it often rests on neural-vocoder artifacts in high-frequency bands above $8$~kHz---precisely the content that GSM and Opus compression discard to conserve bandwidth---so robust artifact-based detection must reason from indirect evidence such as sub-band energy ratios or phase-coherence statistics that overlap with genuine speaker variability.}

\added{Deployment imposes its own demands. On-device detection that avoids cloud offloading calls for a real-time factor below $0.1$ on commodity mobile processors, a target in tension with the transformer-based, self-supervised front-ends that currently generalize best, whose quadratic attention cost resists both quantization and distillation without eroding the very artifacts detection depends on. And because any fixed detector invites evasion, evaluations should routinely include adaptive deepfakes optimized against the defender's own model, so that reported robustness reflects an adversary who adapts rather than a static benchmark.}

\subsection{Adversarial Spoofing Attacks}
\added{Adversarial spoofing is best understood as a dual-objective attack that simultaneously satisfies an impersonation goal against the verifier and an evasion goal against the countermeasure, and defenses that treat these objectives separately are, almost by definition, incomplete. A prerequisite for progress is agreement on what counts as imperceptible: perturbations bounded by $L_\infty < 0.002$ or $L_2 < 0.5$ relative to the waveform offer a starting point, but $\ell_p$ norms are a weak proxy for human hearing, since the same budget may be inaudible when confined to masked frequency bands and obvious otherwise. A defensible protocol will need to couple norm constraints with a psychoacoustic criterion, even though such criteria are non-differentiable and complicate the attacker's optimization in ways that themselves deserve study.}

\added{Strengthening the countermeasures requires more than adversarial training on perturbed examples, though that remains a necessary baseline. Because countermeasures tend to overfit to the spoofing types seen during training, generalization is more likely to come from signal-level fingerprinting or multi-modal biometric fusion than from ever-larger single-modality models, and passive defenses should be complemented by active strategies---randomized challenge prompts or timing jitter---that disrupt the attacker's optimization loop rather than waiting to be probed. Deployment, finally, rewards architectures that decide early: an early-exit defense able to reject adversarial spoofing within the first $500$~ms of audio at $99\%$ precision would be valuable, but is difficult because half a second is shorter than a single prosodic envelope, forcing the decision onto micro-features---vocoder ripples, phase discontinuities---that the attack is explicitly designed to suppress.}

\subsection{Common Challenges Across Threat Families}
\added{Several difficulties recur across all four threat families and, as the critical analysis in Section~\ref{sec:comparison} underscores, arguably matter more than any single attack. The first is transferability: attacks and defenses alike generalize poorly across models, datasets, and languages, leaving universal perturbations on the offensive side and domain-invariant detectors on the defensive side as the most consequential open problems. The second is realism in evaluation, where benchmarks routinely omit compression, ambient noise, spontaneous dialogue, and cross-device variability. Here the community has begun to respond, and ASVspoof~5~\cite{wang2024asvspoof5} is instructive: it assembles crowdsourced speech at scale, incorporates adversarial attacks for the first time, and replaces isolated countermeasure scoring with spoofing-robust automatic speaker verification (SASV) metrics that evaluate the verifier and its countermeasure jointly. Adopting this deployment-oriented, joint evaluation paradigm, in place of the bespoke per-paper protocols that still dominate, would do more to make results comparable than any single methodological advance. The remaining challenges are efficiency, since both attacks and defenses must ultimately run under real-time, low-power, and bandwidth constraints on mobile and edge hardware, and terminology, since fair comparison depends on a shared vocabulary that cleanly separates white-box (gradient), grey-box (score), and black-box (hard-label) access rather than using these terms loosely.}

\subsection{Emerging Attack Paradigms}
\added{Beyond the four families analyzed above, several developments are reshaping the landscape quickly enough that we treat them separately.}

\added{\textbf{Diffusion-based synthesis.} Denoising diffusion probabilistic models~\cite{popov2021grad} now rival or exceed earlier vocoders in perceptual quality, yet their security implications remain largely unstudied. Because a diffusion model generates speech through a stochastic denoising process, it leaves spectral traces qualitatively different from those of GAN-based systems---traces that detectors trained on the ASVspoof lineage were never designed to recognize. A systematic evaluation of diffusion-generated voices against both verifiers and current countermeasures is an urgent and largely unaddressed gap.}

\added{\textbf{Multimodal deepfakes.} The 2024 Arup Engineering fraud showed that attackers have moved from audio-only to synchronized audio-visual deepfakes~\cite{arup2024deepfake}, and the resulting threat model differs in kind: a system that correctly flags audio-domain artifacts may still be defeated when the audio is perceptually reinforced by a convincing video forgery of the same speaker. Joint audio-visual anti-spoofing, and more broadly the study of how cross-modal reinforcement undermines single-modality defenses, is a nascent but increasingly necessary direction.}

\added{\textbf{Large audio-language model security.} As voice authentication is reframed as a language-reasoning task, its language-model component inherits an attack surface with no analog in embedding-based systems. Recent work has already demonstrated stealthy adversarial jailbreaks of large audio-language models~\cite{advwave2024, audiojailbreak2025} and context-agnostic, perceptually imperceptible auditory prompt injection~\cite{audiohijack2024}, and has shown that jailbreaks can transfer across modalities in omni-models, so that a weakness first exposed in text resurfaces through the audio channel. Defenses remain immature, and the field still lacks a shared benchmark for audio-language verification security, without which progress cannot be measured. In our assessment, principled threat modeling for this setting---spanning indirect prompt injection through recorded audio, the robustness of safety alignment, and the interaction between speaker verification and instruction following---is among the most urgent gaps this survey identifies.}

\added{\textbf{Proactive defense and the watermarking arms race.} A structural shift is underway from detecting synthetic speech after the fact to marking it at the moment of generation. Localized audio watermarking~\cite{sanroman2024audioseal} can embed imperceptible, sample-level marks that survive common manipulations and support real-time detection, while provenance standards such as C2PA aim to attach tamper-evident metadata to generated media. For a survey of attacks, the salient point is that these defenses define a new offensive objective: watermark removal, forgery, and evasion are themselves becoming an active research area, and the durability of proactive marking under adaptive removal attacks is far from settled.}

\added{\textbf{Source tracing and forensic attribution.} Detection establishes only that audio is synthetic; forensic and legal use increasingly demands attribution---identifying which system produced a sample and explaining the decision. Source tracing and generator attribution have emerged as distinct tasks~\cite{sourcetracing2024}, with current methods struggling to generalize in the open-set regime where the synthesizer is unknown, and with explainability and partial-spoof localization---determining which segments of an otherwise genuine recording were manipulated---remaining largely open. These capabilities are the natural counterpart to the partial-fake and zero-shot cloning attacks surveyed earlier, and their development will be essential wherever detection outcomes must withstand scrutiny.}

\subsection{Urgent Research Priorities}
\added{Among these many directions, three demand immediate attention because deployed systems are already exposed. The first is real-time deepfake detection at the edge: lightweight models under roughly five megabytes that detect streaming synthesis with latency below $200$~ms and a real-time factor under $0.1$ on standard mobile CPUs, without cloud offloading. This is hard because streaming forces a decision before the full utterance---and with it the long-range prosodic cues that most reliably separate synthetic from natural speech---is available, and no published architecture yet satisfies both the size and latency budgets at once. The second is a genuinely unified defense against optimization-based attacks: rather than maintaining separate pipelines for logical-access spoofing and adversarial perturbation, the field needs a single model that holds equal error rate below five percent against both high-fidelity text-to-speech and $L_\infty$-bounded perturbations at $\epsilon = 0.002$. The third is certifiable robustness. Empirical defenses are broken by adaptive attacks with dispiriting regularity, and only guarantees that predictions remain stable for every perturbation within a defined norm ball---for example $L_2 < 0.5$---will end the cat-and-mouse cycle for bounded threats. Progress on these three fronts, measured under the standardized and realistic protocols described above, is the most direct route from today's reactive posture to systems that can withstand the capabilities the literature has already demonstrated.}

\section{Conclusion}
\label{sec:conculsion}

This survey identifies a fundamental trade-off in contemporary voice authentication systems. While deep learning has substantially improved authentication accuracy and usability, it has also expanded the attack surface and introduced vulnerabilities that are increasingly difficult to characterize, compare, and defend against. Among the threats reviewed, the \textbf{most pressing} is the high-fidelity zero-shot deepfake: it requires no access to the target model, can be produced from only seconds of reference audio, and is convincing enough to deceive both automated verifiers and human listeners, a combination that has already enabled real-world impersonation fraud at scale. The \textbf{most critical defense gap} is the persistent separation between anti-spoofing countermeasures and adversarial robustness. These defenses are typically developed and evaluated in isolation, yet adversarial spoofing attacks show that synthetic speech generation and perturbation-based evasion can be combined, leaving each defense layer vulnerable when the other is also under pressure.

Three priorities follow. First, deepfake detection must move from offline benchmarks toward lightweight, on-device inference, so that verification decisions can be made in real time before the synthetic audio is acted upon. Second, defenses must be unified rather than stacked: a single model should withstand both high-fidelity synthetic speech and bounded adversarial perturbations, since attackers already combine these surfaces in practice. Third, the field must move beyond empirical defenses, which are repeatedly broken by adaptive attacks, toward certifiable robustness that provides mathematical guarantees over bounded perturbation regions. Concrete progress on these fronts, evaluated under standardized and realistic threat models, is the most direct path from the current reactive posture toward voice authentication systems that can withstand the attack capabilities the literature has already demonstrated.

\section*{Acknowledgments}
This research is supported by the Air Force Office of Scientific Research under award number FA2386-23-1-4003.

\section*{Declarations}

\begin{itemize}
\item \textbf{Funding}
\end{itemize}
This work was supported by the Air Force Office of Scientific Research (award number FA2386-23-1-4003).
\begin{itemize}
\item \textbf{Conflict of interest/Competing interests}
\end{itemize}
The authors declare that they have no known competing financial interests or personal relationships that could have appeared to influence the work reported in this paper.

\begin{itemize}
\item \textbf{Declaration of generative AI and AI-assisted technologies in the writing process}
\end{itemize}
During the preparation of this work, the author(s) used ChatGPT 4.0 to check grammar, correct language, and paraphrase certain sentences to improve readability. After using this tool, the author(s) reviewed and edited the content as needed and take full responsibility for the content of the published article.

\begin{itemize}
\item \textbf{Data availability}
\end{itemize}
No data was used for the research described in the article.

\bibliography{sn-bibliography}

\end{document}